\documentclass{aa}  
\usepackage{graphicx}
\usepackage{dsfont}
\usepackage{natbib}
\usepackage{txfonts}
\usepackage{footmisc}
\usepackage[percent]{overpic}
\usepackage[colorlinks=true, allcolors=blue,citecolor=blue,linkcolor=blue]{hyperref}
\usepackage{xcolor}
\newcommand{\ca}{\mbox{Ca\,{\textsc{ii}}~K\,}}

\definecolor{greentheo}{rgb}{0.15,0.50,0.30}

\begin{document} 
\sloppy
   \title{Revisiting the SATIRE-S irradiance reconstruction: Heritage of Mt~Wilson magnetograms and \ca observations} 
   	\titlerunning{SATIRE-S incorporating Mt Wilson magnetograms and \ca data}
   \author{Theodosios Chatzistergos
          \inst{1}
          \and
          Natalie A.  Krivova
           \inst{1}
          \and
           Sami K Solanki
           \inst{1}
           \and
          Kok Leng Yeo
           \inst{1}
          }

   \institute{Max Planck Institute for Solar System Research, Justus-von-Liebig-Weg 3, G\"{o}ttingen, 37077, Germany\\
              \email{\href{mailto:chatzistergos@mps.mpg.de}{chatzistergos@mps.mpg.de}}
             }

  \abstract
{Accurate information on long-term variations in solar irradiance, important for understanding the solar influence on Earth's climate, cannot be derived from direct irradiance measurements due to the comparatively short lifetimes of space-borne experiments.
Models using measurements of the solar photospheric magnetic field as input can provide an independent assessment of the changes.} 
{The Spectral And Total Irradiance Reconstruction in the satellite era (SATIRE-S) model does just that. 
Unfortunately, the magnetogram archives used by  SATIRE-S to recover irradiance variations are also relatively short-lived and have short mutual overlapping periods, making it difficult to evaluate their consistency.
To overcome this and bridge the various archives more reliably, we include additional input data sets.}
{We improve SATIRE-S total solar irradiance (TSI) reconstruction by firstly incorporating magnetograms from the Mt Wilson Observatory as well as  unsigned magnetograms reconstructed from Meudon, Rome, and San Fernando \ca data, and secondly, by re-analysing all periods of overlaps between the various archives.}
{Our combined daily irradiance reconstruction from all eight input archives returns an excellent agreement with direct measurements of irradiance, in particular we find a correlation coefficient of 0.98 when comparing to TSIS1/TIM (Total and Spectral Solar Irradiance Sensor Total Irradiance Monitor) data.
The minimum-to-minimum TSI difference between 1976 and 2019 is -0.2$\pm0.17$~Wm$^{-2}$, while the TSI difference between the 1986 and 2019 minima is statistically insignificant (-0.06$\pm0.13$~Wm$^{-2}$). 
Our analysis also sheds light on the trend shown by the TSI over the so-called ACRIM gap, disfavouring a hypothesised increasing trend in TSI in that period.}
{By including more direct and indirect magnetogram time series, we have made the TSI reconstructed by SATIRE-S more robust and accurate. 
The new series shows a reduced trend of decreasing TSI over the last half century, which agrees well with most composites of measured TSI.}


   \maketitle

\section{Introduction}
The primary source of energy for Earth's system is the radiation of the Sun \citep[see e.g.][]{kren_where_2017}. 
Consequently, any fluctuations in solar irradiance have the potential to impact Earth's climate \citep[e.g.][]{haigh_sun_2007, gray_solar_2010,solanki_solar_2013-1,krivova_solar_2018}. 
Hence, acquiring accurate information about the long-term changes in solar irradiance is crucial for understanding the Sun's impact on Earth's climate \citep{masson-delmotte_climate_2021}.

Solar irradiance represents the solar power received per unit area at a distance of one astronomical unit (AU), normalized to the mean Sun-Earth distance. When irradiance is measured across different wavelengths, it is termed spectral solar irradiance (SSI), while after wavelength integration it is called total solar irradiance (TSI).
Since 1978, various space-borne instruments have recorded the fluctuations in TSI \citep[e.g.][]{frohlich_observations_2000,kopp_magnitudes_2016}. 
However, the individual missions covered relatively short periods of time, mostly shorter than one solar cycle, with the Variability of Irradiance and Gravity Oscillations \citep[VIRGO;][]{frohlich_virgo_1995} onboard the Solar and Heliospheric Observatory \citep[SoHO;][]{domingo_soho_1995} being the currently longest one (operating since 1996).

Consequently, data from various missions have been amalgamated into composite series. 
Each instrument features a different absolute scale, noise level, sensitivity changes, and instrumental degradation, which significantly
complicates assessments of any secular variation from direct measurements.
Three commonly used, ``classical'', TSI composites are called ACRIM\footnote{Available at \url{https://web.archive.org/web/20170611210135/http://acrim.com/}} \citep[Active Cavity Radiometer Irradiance Monitor, which is the instrument taken as the reference by][]{willson_total_1997,willson_composite_2003}, PMOD\footnote{Available at \url{https://www.pmodwrc.ch}} \citep[named after Physikalisch-Meteorologisches Observatorium Davos;][]{frohlich_solar_2006}, and ROB\footnote{Available at \url{https://www.sidc.be/observations/space-based-timelines/tsi}} \citep[named after Royal Observatory of Belgium, previously referred to as RMIB, Royal Meteorological Institute of Belgium;][]{dewitte_total_2004,dewitte_total_2016}. 
More recently, a few more composites have emerged, such as those by \citet{dudok_de_wit_methodology_2017}\footnote{More recent versions are available at \url{https://spot.colorado.edu/~koppg/TSI/}}, \citet{schmutz_changes_2021}, \citet[][Composite PMOD-Data Fusion, CPMDF, hereafter]{montillet_data_2022}, and Copernicus Climate Change Service (C3S)\footnote{Available at: \url{https://confluence.ecmwf.int/pages/viewpage.action?pageId=304239361}}. 
These composites exhibit partly diverging long-term trends, usually characterised by the changes between irradiance levels during different cycle minima \citep{frohlich_total_2012,kopp_magnitudes_2016,chatzistergos_long-term_2023}. 
While all published composites evidence a decrease in TSI since 1996, they disagree regarding the exact magnitude and the direction of the trend between 1986 and 1996. 
The persisting uncertainty precludes definitive conclusions regarding secular changes based only on direct TSI measurements \citep{montillet_data_2022}.

By now it is established that variability of solar irradiance stems from changes in the solar surface magnetic field  \citep{shapiro_nature_2017,yeo_solar_2017}. 
Thus,  the most straightforward way to reconstruct irradiance variations is by using directly observed maps of the solar surface magnetic field.
This is the approach taken by the
SATIRE \citep[Spectral And Total Irradiance REconstruction;][]{fligge_modelling_2000-1,krivova_reconstruction_2003} model. 
Various versions of SATIRE exist, depending on the input data and covered period. 
SATIRE-S, where ``S'' stands for the satellite era \citep{krivova_reconstruction_2003,wenzler_reconstruction_2006,ball_reconstruction_2012,yeo_reconstruction_2014} uses various space- and ground- based magnetograms to reconstruct irradiance back to 1974. 
The data from the different datasets were merged by cross-calibrating them with the histogram equalisation method \citep{jones_preliminary_2001}. 
In particular, the most recent version of SATIRE-S by \citet{yeo_reconstruction_2014} included the data from the Helioseismic and Magnetic Imager \citep[HMI;][]{scherrer_helioseismic_2012} onboard the space-based Solar Dynamics Observatory \citep[SDO;][]{pesnell_solar_2012}, the Michelson Doppler Imager \citep[MDI;][]{scherrer_solar_1995} on board the space-based SoHO, and the Kitt Peak Vacuum Telescope using a 512-channel Babcock type instrument (KP/512) and a CCD spectromagnetograph (KP/SPM).
The reconstruction by \citet{yeo_reconstruction_2014} resulted in a more pronounced decline between the activity minima of 1986 and 1996 compared to the composites of direct measurements.
\citet{yeo_variation_2024} have shown that the stability of the instrument response was not an issue and could not be responsible for such a trend.
This could, however, still potentially be due to  residual issues with cross-calibration of the individual datasets.
The individual magnetogram archives used in that reconstruction covered roughly one solar cycle each, while the overlap between them was not always optimum to allow an accurate calibration, which could introduce uncertainty in the resulting long-term trend.
For instance, the cross-calibration of the KP/512 and KP/SPM was based on just 11 days of joint observations. 

In this work, we make use of four additional datasets to better understand the potential effect of the instrumental switches on the long-term trend of the SATIRE-S reconstruction. 
These include directly recorded magnetograms from the Mt Wilson Observatory and three sets of unsigned magnetograms reconstructed from \ca observations. These additional datasets allow creating a more reliable reconstructed composite irradiance dataset.

We briefly describe the SATIRE-S model in Section~\ref{sec:satires}, followed by an overview of the data we used and their processing in Section~\ref{sec:data}.
We outline the overall process to reconstruct irradiance and present the new SATIRE-S composite reconstruction in Section~\ref{sec:tsirecs}.
In Section~\ref{sec:compmeasurements} we compare our new TSI and SSI reconstructions to irradiance measurements and discuss the activity minimum-to-minimum trend of the final series.
Finally, we summarise our results in Section~\ref{sec:summary}.

\section{The SATIRE-S model}
\label{sec:satires}

The model has been extensively described in the literature \citep{fligge_modelling_2000-1,krivova_reconstruction_2003,wenzler_reconstruction_2006,ball_reconstruction_2012,yeo_reconstruction_2014}, thus here we only give a brief overview.

SATIRE-S divides the solar surface into four different components, faculae (f), sunspot umbra (u), sunspot penumbra (p), and quiet Sun (q).
The intensities $I_{\mathrm{u,p,f,q}}(\mu;\,\lambda)$ of each component as a function of wavelength, $\lambda$, and position on the disc in terms of $\mu$, the cosine of the heliocentric angle, were computed with the radiative transfer code ATLAS9 \citep{kurucz_atlas_1970} by \citet{unruh_spectral_1999} from the appropriate semi-empirical model atmospheres. 
The quiet Sun is represented by the \citet{kurucz_new_1993,kurucz_atlas12_2005} model at T = 5777 K, sunspot umbrae and penumbrae by the models with T = 4500 K and T = 5450 K, respectively, and faculae and network by the FALP model \citep{fontenla_calculation_1999} as modified by \citet{unruh_spectral_1999}.
The synthesized spectra range from 115 to 160,000~nm.
The wavelength grid of the spectra, which is also the same for our SSI reconstruction, is described in Table \ref{tab:wlgrid}.

\begin{table}
\caption{Wavelength grid for the synthesised spectra.}
\label{tab:wlgrid} 
\centering
\begin{tabular}{lc}
\hline\hline
Central wavelength [nm]&step [nm]\\
\hline                    
\centering
115.5--289.5  	& 1\\
291--999	 	& 2\\
1,002.5--1,597.5& 5\\
1,605--3,195	&10\\
3,210--6,390	&20\\
6,420--10,020	&40\\
20,000--160,000	&20,000\\
\hline
\end{tabular}
\end{table}

To derive the distribution of faculae and sunspots on the solar visible surface, we use full-disc magnetograms and continuum observations, respectively.
For this, we analyse every pixel of each image to determine whether it belongs to sunspot umbra, penumbra, facula or quiet Sun.
Sunspots are identified as pixels with continuum contrasts below certain thresholds \citep[i.e. large negative contrasts; see][and Sect. \ref{sec:fillingfactors} for details]{yeo_reconstruction_2014}. 
All other pixels are classified based on their magnetogram signal. 
If the magnetic flux density of a pixel is below the noise threshold \citep[see][and Sect. \ref{sec:fillingfactors}]{yeo_reconstruction_2014}, they are considered to be free of a detectable magnetic field and are thus attributed to the quiet Sun. 
All other pixels are assumed to lie in facular regions.
After labeling each pixel as belonging to umbra, penumbra, faculae, or quiet Sun, we calculate the filling factors as the disc fractional area of each component across 101 equidistant $\mu$ rings.

For sunspots (including both umbrae and penumbrae), the labeling process is binary, in the sense that each pixel is assigned a value of one if it corresponds to that component or zero if it does not.
For faculae, however, the relationship between magnetic flux, as measured in magnetograms, and the actual brightness of the corresponding magnetic regions on the Sun is more uncertain \citep[see, e.g.,][]{fligge_modelling_2000, yeo_reconstruction_2014}.
To address this, we consider $A_{\mathrm{f}}(i,j)$ which expresses the area fraction of pixel $i,j$ that is covered by faculae.
$A_{\mathrm{f}}(i,j)$ is scaled linearly with the measured pixel magnetic flux density until a saturation limit, $B_{\mathrm{sat}}$, is reached:
\begin{align}
    A_{\mathrm{f}}(i,j) = 
    \begin{cases} 
    |B_r(i,j)| / B_{\mathrm{sat}}\;
    & \mathrm{if} \; |B_r(i,j)| < B_{\mathrm{sat}}\\
    \;\;1\;&\mathrm{if}  \;|B_r(i,j)| \geq B_{\mathrm{sat}}\\
    \end{cases},
   \label{eq:bsat}
\end{align}
where $|B_r(\mu)|$ is the unsigned radial magnetic field strength. 
It is derived by dividing the measured flux density by $\mu$, the cosine of the heliocentric angle, to account for foreshortening. 
The saturation of the facular brightness above a certain threshold is a simple way of accounting for the lower average brightness of magnetic features in regions with more magnetic flux \citep[e.g.][]{solanki_properties_1984} and of the non-linear dependence of brightness of as a function of magnetic flux \citep[e.g.][]{kahil_brightness_2017}.

Then the filling factors of faculae, $\alpha_{\mathrm{f}}(\mu)$, are defined as: 
\begin{equation}
	\alpha_{\mathrm{f}}(\mu_k) = \frac{\sum_i \sum_j A_{\mathrm{f}}(i,j) \mathds{1}_{\left( \mu_{k+1} < \mu_{i,j} \le \mu_k\right)}}{A},
	\label{eq:aff}
\end{equation}
where $\mu_k$ are the central locations of 101 equidistant $\mu$-bins with an index $k$ ranging from 1 to 101, $\mathds{1}$ is the indicator function getting the value of one or zero depending on whether the criterion in the parenthesis is fulfilled, and $A$ is the total number of pixels of the solar disc.
The saturation limit, $B_{\mathrm{sat}}$, is the sole free parameter of the model.
It is set by comparing the TSI reconstructions to direct measurements and is treated as a constant throughout a given magnetogram archive (see Sect. \ref{sec:fixbsat}).

Thus, the solar spectral irradiance, $F(\lambda, t)$, is then computed as:
   \begin{align}
    F(\lambda, t) = \sum_{\mu} 
    \alpha_\mathrm{u}(\mu;t) I_\mathrm{u}(\mu;\,\lambda) +
    \alpha_\mathrm{p}(\mu;t) I_\mathrm{p}(\mu;\,\lambda)\nonumber\\ + 
    \alpha_\mathrm{f}(\mu;t) I_\mathrm{f}(\mu;\,\lambda) +  
    \alpha_\mathrm{q}(\mu;t) I_\mathrm{q}(\mu;\,\lambda).
   \label{eq:rec}
   \end{align}
TSI is then computed by simply integrating $ F(\lambda, t)$ over all wavelengths.

\begin{figure*}
    \centering
\includegraphics[width=1\linewidth]{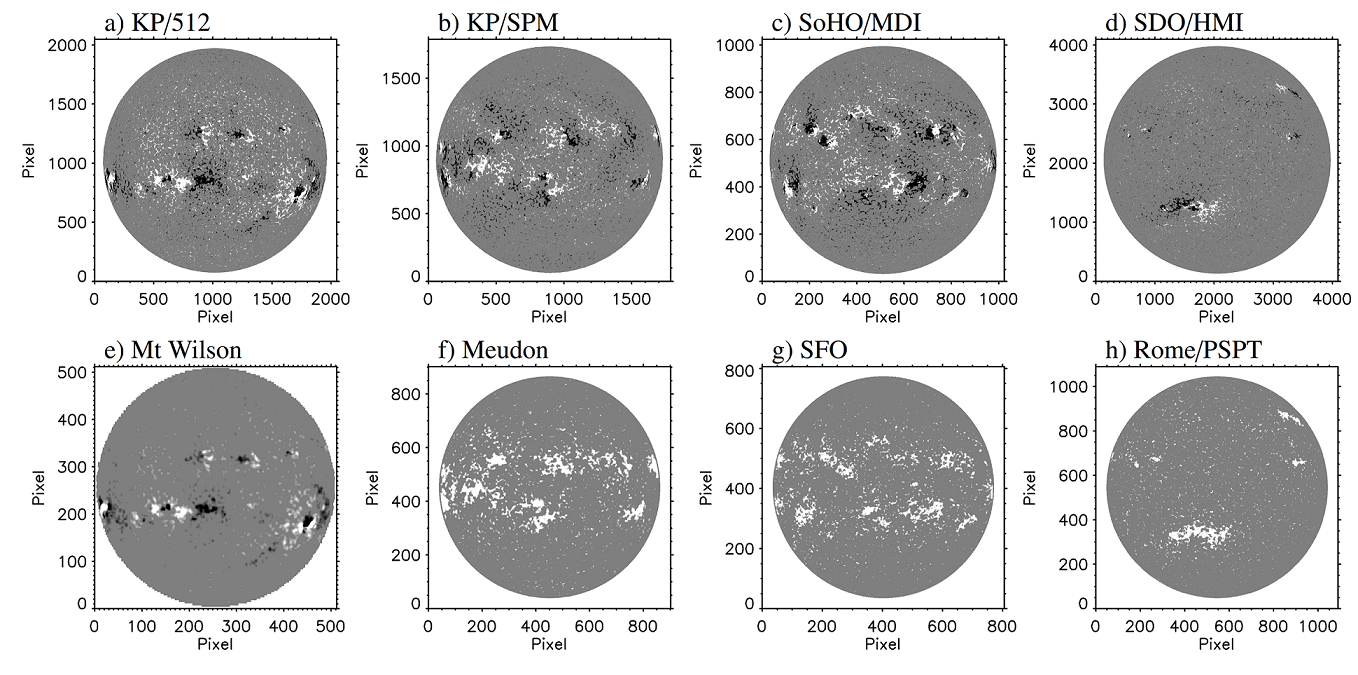}
    \caption{Examples of magnetograms from the various archives used in this study.     
    The dates of the observations from left to right are 30 May 1983, 28 April 1993, 23 April 2002, and 10 November 2010. All magnetograms are saturated at $\pm$ 50 G after removing the signal below the respective noise levels. The three images obtained from \ca observations correspond to unsigned magnetic field.}
    \label{fig:magnetograms}
\end{figure*}

\section{Data and processing}
\label{sec:data}

\begin{table*}
\caption{Characteristics of magnetogram archives used in this study.}
\label{tab:mwdatachains} 
\scriptsize
\centering
\begin{tabular}{l*{11}{c}}
\hline\hline
Dataset&Platform&Type&Period&\multicolumn{2}{c}{Noise level}&Pixel scale&Sunspot source& N images & \multicolumn{2}{c}{N days}&Chained to\\
&&&&Mean&$\sigma$&[''/pixel]&&& Total & Composite&\\

\hline                    
\centering
KP/512   & ground-based & magnetogram & 1974--1993 &8.0& 1.2&1.0& FD continuum                &  1371 & 1371 & 1360 & Meudon\\
Meudon   & ground-based & \ca         & 1974--2017 &\multicolumn{2}{c}{-}&0.9--2.2\tablefootmark{a}& \citet{mandal_sunspot_2020} & 12262 & 9863 & 4124 & SoHO/MDI\\
MW I     & ground-based & magnetogram & 1983--1988 &2.0& 0.3&3.8& \citet{mandal_sunspot_2020} &  8911 & 1476 &  352 & Meudon\\
MW II    & ground-based & magnetogram & 1988--1995 &2.0& 0.3&3.8& \citet{mandal_sunspot_2020} & 18482 & 2049 & 398  & Meudon\\
MW III   & ground-based & magnetogram & 1996--2013 &2.0& 0.3&5.7& \citet{mandal_sunspot_2020} & 45180 & 3344 &  50  & SoHO/MDI\\
KP/SPM   & ground-based & magnetogram & 1992--2003 &4.0& 0.7&1.1& FD continuum                &  2055 & 2055 & 1145 & Meudon\\
SFO      & ground-based & \ca         & 1992--2013 &\multicolumn{2}{c}{-}&2.6& \citet{mandal_sunspot_2020} &  4004 & 3856 &  191 & SoHO/MDI\\
SoHO/MDI & space-based  & magnetogram & 1999--2010 &11.8 &4.0&2.0& FD continuum                &  3941 & 3941 & 3745 & SDO/HMI\\
Rome/PSPT& ground-based & \ca         & 2000--2022 &\multicolumn{2}{c}{-}&2.0& FD continuum &  3206 & 3206 &  133 & SoHO/MDI\\
SDO/HMI  & space-based  & magnetogram & 2010-      &6.4&0.9&0.5& FD continuum                &  5429 & 5429 & 5429 & - \\
\hline
\end{tabular}
\tablefoot{The columns are: name of dataset, ground- or space-based, direct or reconstructed (from \ca) magnetograms, the period covered by the data (as used in this study), mean and standard deviation of noise level, pixel scale, source of the sunspot information (that is from the (quasi-)simultaneous full-disc, FD, continuum data or from the \citealt{mandal_sunspot_2020} database), the total number of images, the number of days that we analysed, the number of days that entered the SATIRE-S composite, and the bridging dataset that was used to cross-calibrate the filling factors to the SDO/HMI level.
\tablefoottext{a}{The CCD-based data since 2002 have a pixel scale of 1.5'', while the photographic data derive from two digitisations with pixel scales of 0.9'' and 2.2''.}
}

\end{table*}

\subsection{Full-disc observations }
\label{sec:mag}
To extract the distribution of the different magnetic features over the solar surface and its changes with time,
in this work we use various space- and ground-based full-disc observations of the Sun, including magnetograms, 
continuum images, as well as images in the Ca~II~K line converted into unsigned magnetograms.
Table \ref{tab:mwdatachains} gives an overview of the various individual archives, while below we briefly describe them. 
The most recent published version of SATIRE-S \citep{yeo_reconstruction_2014} used the following datasets: Helioseismic and Magnetic Imager \citep[HMI;][]{scherrer_helioseismic_2012} onboard the space-based Solar Dynamics Observatory \citep[SDO;][]{pesnell_solar_2012}, Michelson Doppler Imager \citep[MDI;][]{scherrer_solar_1995} on board the space-based SoHO, and Kitt Peak National Solar Observatory (Sects. \ref{sec:data_HMI}--\ref{sec:data_Kittpeak}). Here, we additionally include magnetograms from Mt Wilson (Sect. \ref{sec:data_mtwilson}) and the unsigned magnetograms reconstructed from \ca data. 
\ca brightness is known to be an excellent proxy of solar surface magnetism \citep{babcock_suns_1955,skumanich_statistical_1975,schrijver_relations_1989,loukitcheva_relationship_2009,kahil_brightness_2017,chatzistergos_recovering_2019}. 
Thus, it is possible to use \ca observations to reconstruct unsigned magnetograms \citep{chatzistergos_recovering_2019}. 
In particular, \citet{chatzistergos_reconstructing_2021-1} showed that one can reconstruct solar irradiance in this way nearly as well as when using directly recorded magnetograms. 
Therefore, here we also make use of three datasets of unsigned magnetograms reconstructed from \ca images.  
The three most useful archives for our purpose are those from Rome, San Fernando (SFO), and Meudon.

\subsubsection{SDO/HMI}
\label{sec:data_HMI}
SDO/HMI started operation on 30th of April 2010 and continues to this day.
It continuously captures full-disc filtergrams at six wavelength positions along the Fe I 6173 \AA~ line. 
These filtergram data are merged to generate simultaneous continuum intensity images and longitudinal magnetograms at a cadence of 45 seconds. 
Following \citet{yeo_reconstruction_2014} we keep only one daily intensity image and magnetogram calculated as the average over a 315-second period in order to mitigate signal fluctuations from noise and p-mode oscillations.
As \citet{yeo_reconstruction_2014}, we also chose the data closest in time to 12 UT, except for the period of overlap to SoHO/MDI observations (April 30 to December 24, 2010), for which we selected the SDO/HMI observations closest in time to the SoHO/MDI ones.

\subsubsection{SoHO/MDI}
\label{sec:data_MDI}
SoHO/MDI provided observations from March 19, 1996 to April 11, 2011. 
It captured full-disc filtergrams at four wavelength positions along the Ni I 6768 \AA~ line and additionally at one wavelength in the nearby continuum, generating continuum intensity images and longitudinal magnetograms on each observation day. 
Continuum images and magnetograms are, however, not synchronous. 
Following \citet{ball_reconstruction_2012} and \citet{yeo_reconstruction_2014}, for each day we chose the level 1.5 continuum intensity image and the level 1.8.2 5-minute longitudinal magnetogram \citep{liu_comparison_2012} recorded closest in time to each other. 
However, as was also done by \citet{ball_reconstruction_2012} and \citet{yeo_reconstruction_2014} we excluded SoHO/MDI observations prior to February 2, 1999, due to changes in the instrument response to magnetic flux during extended outages suffered by the SoHO spacecraft between June 1998 and February 1999, the so-called SoHO-vacation \citep[see also][]{yeo_variation_2024}. 
Following \citet{yeo_reconstruction_2014}, we also excluded the SoHO/MDI data after December 24, 2010 due to issues with the flat field of the observations.

\subsubsection{Kitt Peak}
\label{sec:data_Kittpeak}
Daily full-disc continuum intensity images and longitudinal magnetograms based on spectropolarimetry of the Fe I 8688 \AA~line were obtained with the ground-based Kitt Peak Vacuum Telescope using a 512-channel Babcock type instrument (KP/512) from February 1, 1974, to April 10, 1993. 
Additionally, data were collected using a CCD spectromagnetograph (KP/SPM) from November 19, 1992, to September 21, 2003.

We used only  1371 and 2055 pairs of continuum observations and magnetograms from KP/512 and KP/SPM, respectively.
These were selected by \citet{wenzler_reconstruction_2006} and \citet{yeo_reconstruction_2014} out of the total of 4665 and 2894 KP/512 and KP/SPM images, respectively, as those reasonably devoid of instrumental artefacts and not unduly affected by atmospheric seeing.

\subsubsection{Mt Wilson}
\label{sec:data_mtwilson}
Mt Wilson observatory started producing photospheric magnetograms in 1967 and continued until 2013,
thus it currently has the longest collection of solar surface magnetograms.
A Babcock solar magnetograph was employed at Mt Wilson, which recorded the Zeeman polarization in the wings of the Fe I 5250 \AA~absorption line \citep{ulrich_analysis_1992}. 
We use the data as updated by \cite{ulrich_calibration_2019,ulrich_calibration_2024}\footnote{Available at \url{ftp://howard.astro.ucla.edu/pub/obs/fits/} and \url{https://dataverse.harvard.edu/dataset.xhtml?persistentId=doi:10.7910/DVN/4VTNJ3}}. 
Unfortunately, various instrumental changes and equipment failures occurred over the years \citep[see e.g.][]{howard_mount_1983,ulrich_system_1991,ulrich_mount_2002,ulrich_calibration_2019,ulrich_calibration_2024} introducing inconsistencies in the series. 
Particularly important for our purpose are the changes made to the instrument in 1981 \citep{howard_mount_1983}, 1988 \citep{ulrich_mount_2002}, and 1996\footnote{\url{https://web.archive.org/web/20230703170118/http://obs.astro.ucla.edu/150_magn.html}}.
The whole magnetograph was largely rebuilt in 1981 \citep{howard_mount_1983}, as earlier data had stability and calibration issues.  
In May 1982, as reported by \citet{howard_mount_1983}, also the grating system was replaced.
In 1988 and 1996 additional fiber-optic reformators were introduced to the system \citep{ulrich_mount_2002}.
In 1996 the instrument was further upgraded into a 24-channel spectral sampling system \citep{ulrich_mount_2002}.

The magnetograms have a pixel scale of $\sim$3.8'' and $\sim$5.7'' before and after 1996, respectively.
Multiple images exist for each day of observation, with the whole dataset (1967--2013) comprising 73263 images. 
For the time being, for our irradiance reconstruction with Mt Wilson data we did not consider the data before 1983 due to uncertainties with their calibration. 
We also removed 450 problematic magnetograms from our analysis.
These were magnetograms which either had parts of the solar disc missing or exhibited a standard deviation less than 1~G so that their values were lower than the average noise level of this dataset.
Furthermore, we considered the rest of the data as three separate sets covering the periods January 1983--July 1988 (MW I), August 1988--December 1995 (MW II), and January 1996-- January 2013  (MW III) (see Sect. \ref{sec:MWtsirecs} for further details).
Table \ref{tab:mwdatachains} lists the relevant information for each of these subsets of Mt Wilson magnetograms.

\subsubsection{Rome/PSPT}
\label{sec:data_PSPT}
Rome is one of the most accurate CCD-based \ca archives with observations available since 1996 \citep{ermolli_rome_2022}. 
The Rome archive contains observations in the \ca line and four continuum intervals with the Precision Solar Photometric Telescope (PSPT) utilising interference filters \citep{ermolli_rome_2022}.
Here, we restrict the use of Rome/PSPT \ca data and red continuum to the period since 2000 to avoid potential issues due to some early instrumental changes \citep{ermolli_rome_2022}.

\subsubsection{San Fernando}
SFO is the longest-running \ca archive producing digital images with a linear diode array extending back to 1988 \citep{chapman_tsi_2024}.
SFO data are taken with two different telescopes, Cartesian Full Disk Telescope no. 1 (CFDT1) and  no. 2 (CFDT2) using an interference filter  \citep{chapman_modeling_2013}.
Here, we only used the observations from CFDT2 since the beginning of 1993 due to their better quality over the CFDT1 ones.
Although continuum observations were performed at SFO, we do not have access to them and thus employ the \citet{mandal_sunspot_2020} sunspot database to derive information about sunspots.

\subsubsection{Meudon}
Meudon is the longest running archive of \ca observations extending back to 1893 (although systematic observations started in 1908) and continuing to the present \citep{malherbe_130_2023,chatzistergos_full-disc_2022}. 
The quality of Meudon data over the period of interest here (since 1974) is sufficiently good for our purpose \citep[see][]{chatzistergos_reconstructing_2021-1,chatzistergos_understanding_2024}. 
Meudon observations were performed with a spectroheliograph and they were stored on photographic plates until 2002, while a CCD camera was used afterwards.
In 2017 the system was upgraded \citep{malherbe_new_2019} which resulted in a change of the bandwidth of the observations.
For this reason in this work we restrict the use of Meudon data to the period before 2017.
Neither do we consider the data prior to 1974, since at this point significantly more work is required to ascertain the consistency of earlier data with the more recent ones.
Unfortunately, continuum observations co-temporal to the Meudon \ca data are not available, thus we use again the \citet{mandal_sunspot_2020} database.

\subsection{Processing}
\label{sec:processing}
We used the SoHO/MDI and Kitt Peak magnetograms processed by \citet{yeo_reconstruction_2014}, who applied histogram equalisation on them and rescaled their magnetogram signal to the SDO/HMI level.
Specifically, first the SoHO/MDI magnetograms were calibrated to the SDO/HMI level, and then subsequently, the Kitt Peak magnetograms were calibrated to the SoHO/MDI, ensuring both were ultimately rescaled to the SDO/HMI level.

The photographic Meudon \ca data were photometrically calibrated (to account for the non-linear response of the photographic plates) with the method by \citet{chatzistergos_analysis_2018,chatzistergos_analysis_2019}.
Unsigned magnetograms were reconstructed from all \ca data used here by applying the relationship derived by \citet{chatzistergos_recovering_2019,chatzistergos_reconstructing_2021-1}.
Since the relationship requires \ca observations corrected for limb darkening, we employed the method developed by \citet{chatzistergos_analysis_2018,chatzistergos_analysis_2020} for this purpose.
The unsigned magnetograms reconstructed from Meudon, SFO, and Rome/PSPT were already, by design, directly at the SDO/HMI level since the relationship we had used to produce them was derived using SDO/HMI data as the reference.

Next, we determined the noise level of the magnetograms (except the maps derived from \ca archives).
For this, we followed \citet{wenzler_reconstruction_2006} and \citet{yeo_intensity_2013}. 
In contrast to \citet{yeo_intensity_2013}, however, we did not use the highest noise level (that is the one of SoHO/MDI) for all archives but derived the noise level for each instrument separately.
Pixels with values lower than $3\sigma$ of the noise level were identified as noise and set to 0~G.
That is they are considered as quiet Sun.
Figure \ref{fig:magnetograms} shows examples of SDO/HMI, SoHO/MDI, KP/SPM, KP/512, and Mt Wilson magnetograms after accounting for their noise levels.
We converted all magnetograms to radial field unsigned maps by dividing them by   $\mu$, the cosine of the heliocentric angle $\theta$, and taking the absolute values.
This last step was not needed for the reconstructed magnetograms from \ca data since they are already in radial field unsigned values.

The continuum observations from Kitt Peak, SoHO/MDI, and SDO/HMI datasets were processed to compensate for the limb darkening as in \citet{yeo_reconstruction_2014}, while red-continuum Rome/PSPT data were corrected using the method by \citet{chatzistergos_analysis_2018,chatzistergos_modelling_2020}.

\subsection{Irradiance data}
\label{sec:tsimeas}
For comparison purposes we used various published TSI series, including measurements of individual instruments, composites of direct measurements, as well as modelled series.

In particular, we used direct measurements from:
\begin{itemize}
    \item Earth Radiation Budget onboard the Nimbus-7 satellite \citep[Nimbus-7/ERB;][covering 1978--1993]{hoyt_nimbus_1992};
    \item SoHO/VIRGO \citep[][covering 1996--2024; version 8]{frohlich_virgo_1995};
    \item Total Irradiance Monitor (TIM) onboard the Solar Radiation and Climate Experiment \citep[SORCE/TIM;][covering 2003--2020; version 19]{kopp_total_2005};
    \item TSI Continuity Transfer Experiment TIM \citep[TCTE/TIM;][covering 2013--2019; version 4]{kopp_tcte_2013};
    \item The Total and Spectral Solar Irradiance Sensor TIM \citep[TSIS1/TIM;][covering 2018--2024; version 4]{pilewskie_tsis-1_2018}.
\end{itemize} 
The data from SoHO/VIRGO, SORCE/TIM, TCTE/TIM, and TSIS1/TIM, which all have direct overlap to SDO/HMI magnetograms, were also used to set $B_\mathrm{sat}$.

We also used the TSI composites by \citet{montillet_data_2022}, ACRIM \citep[][]{willson_total_1997,willson_composite_2003}, ROB \citep{dewitte_total_2016}, \citet{dudok_de_wit_methodology_2017}, and C3S. 
We note, however, that while all composites used here receive regular updates, ACRIM has not received any official update since 2013.
For the \citet{dudok_de_wit_methodology_2017} TSI composite we use the version from 31/10/2016. 
Finally, we also compared our results to the earlier SATIRE-S TSI reconstruction by \citet{yeo_reconstruction_2014}, which we extended to the present.

\subsection{Activity indices}
\label{Sec:proxies}
Even when using partly overlapping magnetograms from all the available archives, there are still some gaps in daily data, especially during the earlier period. To fill those gaps, we follow \citet{yeo_reconstruction_2014} and use additional solar magnetic activity indices: 
\begin{enumerate}
    \item The projected sunspot area composite by \citet{mandal_sunspot_2020}\footnote{\label{sunclimatewebsite}Available at: \url{https://www2.mps.mpg.de/projects/sun-climate/data.html}};
\item The \ca plage projected area composite by \citet{chatzistergos_analysis_2020,chatzistergos_understanding_2024}\footref{sunclimatewebsite};
\item The Lyman-$\alpha$ composite by \citet{machol_improved_2019}\footnote{Available at: \url{https://lasp.colorado.edu/lisird/data/composite_lyman_alpha}};
\item The Ottawa and Penticton adjusted 10.7 cm radio flux \citep{tapping_107_2013}\footnote{Available at:  \url{https://lasp.colorado.edu/lisird/data/penticton_radio_flux}};
\item The Institute of Environmental Physics (IUP), University of Bremen Mg II index \citep{viereck_mg_2001}\footnote{Available at:  \url{https://lasp.colorado.edu/lisird/data/bremen_composite_mgii}}.
\end{enumerate}

\begin{figure*}[]
	\centering
\includegraphics[width=0.8\linewidth]{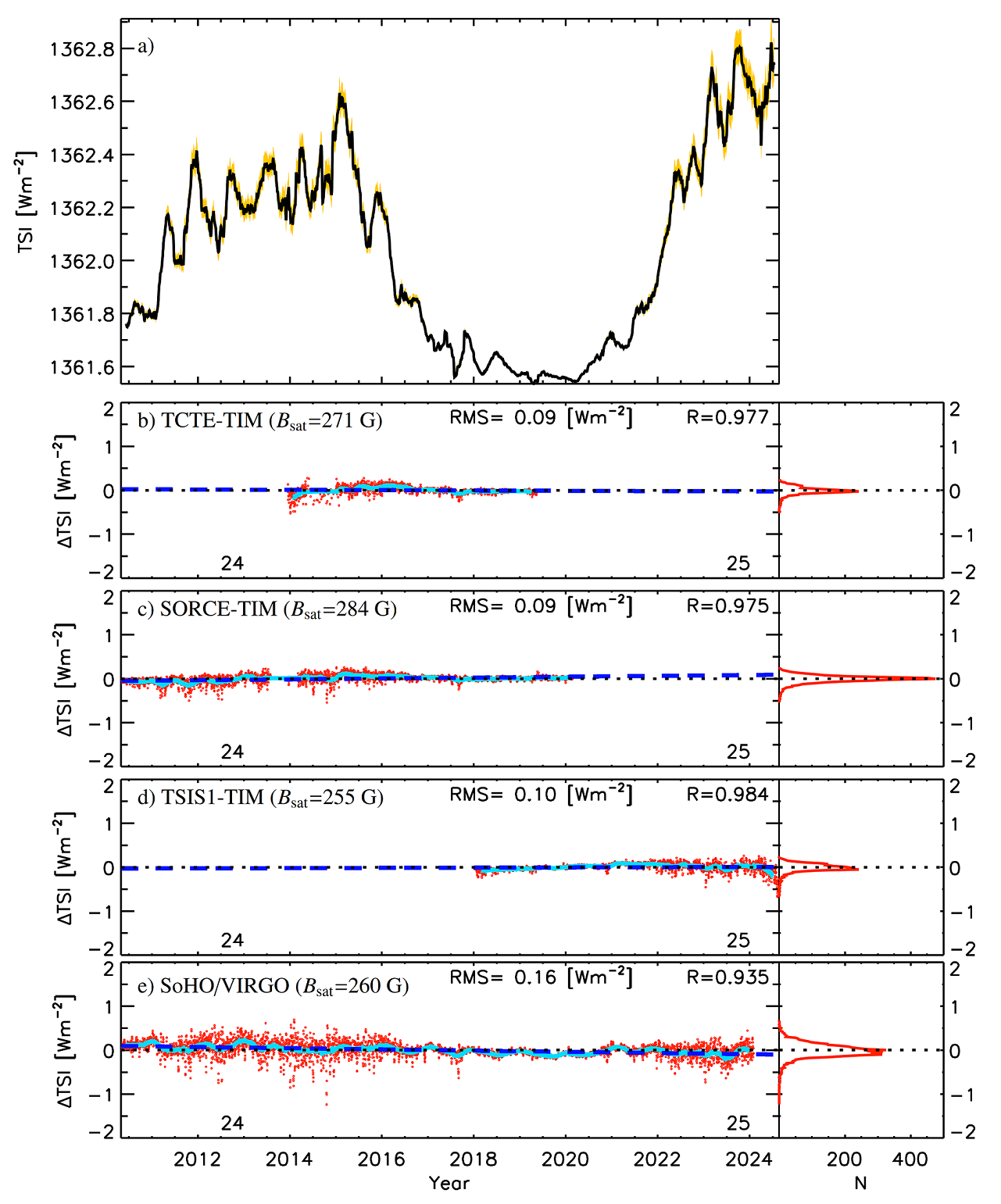}
 	  \caption{\textit{Panel a): }TSI reconstructions with SATIRE-S using SDO/HMI magnetograms. Our adopted reconstruction is shown as 81-day running means in black, while the shaded yellow surface denotes the uncertainty due to the choice of the reference TSI dataset used to set $B_{\mathrm{sat}}$.    \textit{Panels b)--e): } Difference between the TSI reconstructed with the SATIRE-S model using SDO/HMI magnetograms and various TSI series as denoted by the legends. Daily values are represented by red circles, 180-day running means by the ciel curves. All series have been offset to match the value of SORCE/TIM over 2019. The horizontal dotted black line marks residuals of 0 Wm$^{-2}$.  
   The dashed blue lines are linear fits to the residuals.
   Also given within each panel are the RMS differences and linear correlation coefficient, $R$.
   The right panel shows the histogram of the residuals in bins of 0.02 Wm$^{-2}$. }\label{fig:hmi}
\end{figure*}

\begin{figure*}[!htb]
	\centering
\includegraphics[width=0.8\linewidth]{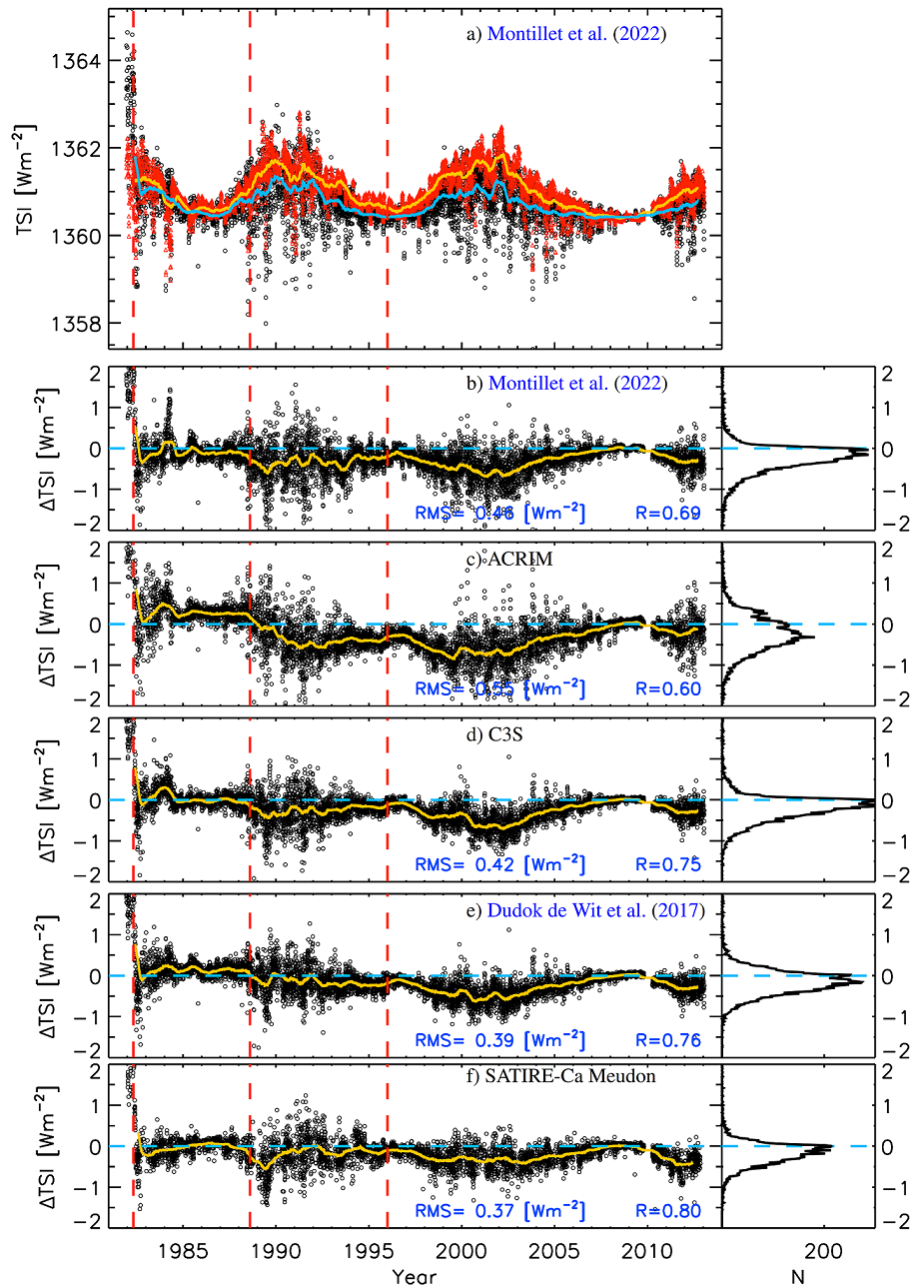}
	  \caption{\textit{panel a): }TSI reconstruction with the SATIRE-S model using Mt Wilson magnetograms (black circles for daily and ciel line for 180-day running means) along with the \citet{montillet_data_2022} TSI composite (red triangles for daily and yellow line for 180-day running means). \textit{Panels b)--f): } Difference between TSI reconstructed with the SATIRE-S model using Mt Wilson magnetograms and various TSI series as denoted by the legends. Daily values are shown as black circles and 180-day running means by yellow curve. All series have been offset to match the value of \citet{montillet_data_2022} TSI composite over 2008. The horizontal dashed ciel line marks residuals of 0 Wm$^{-2}$.  
   The vertical dashed red lines mark periods of relevant instrumental changes in the Mt Wilson data.
   The right panel shows the histogram of the residuals in bins of 0.02 Wm$^{-2}$. 
	  }\label{fig:tsirecmw2_unc}
\end{figure*}

\begin{figure*}[]
	\centering
\includegraphics[width=0.7\linewidth]{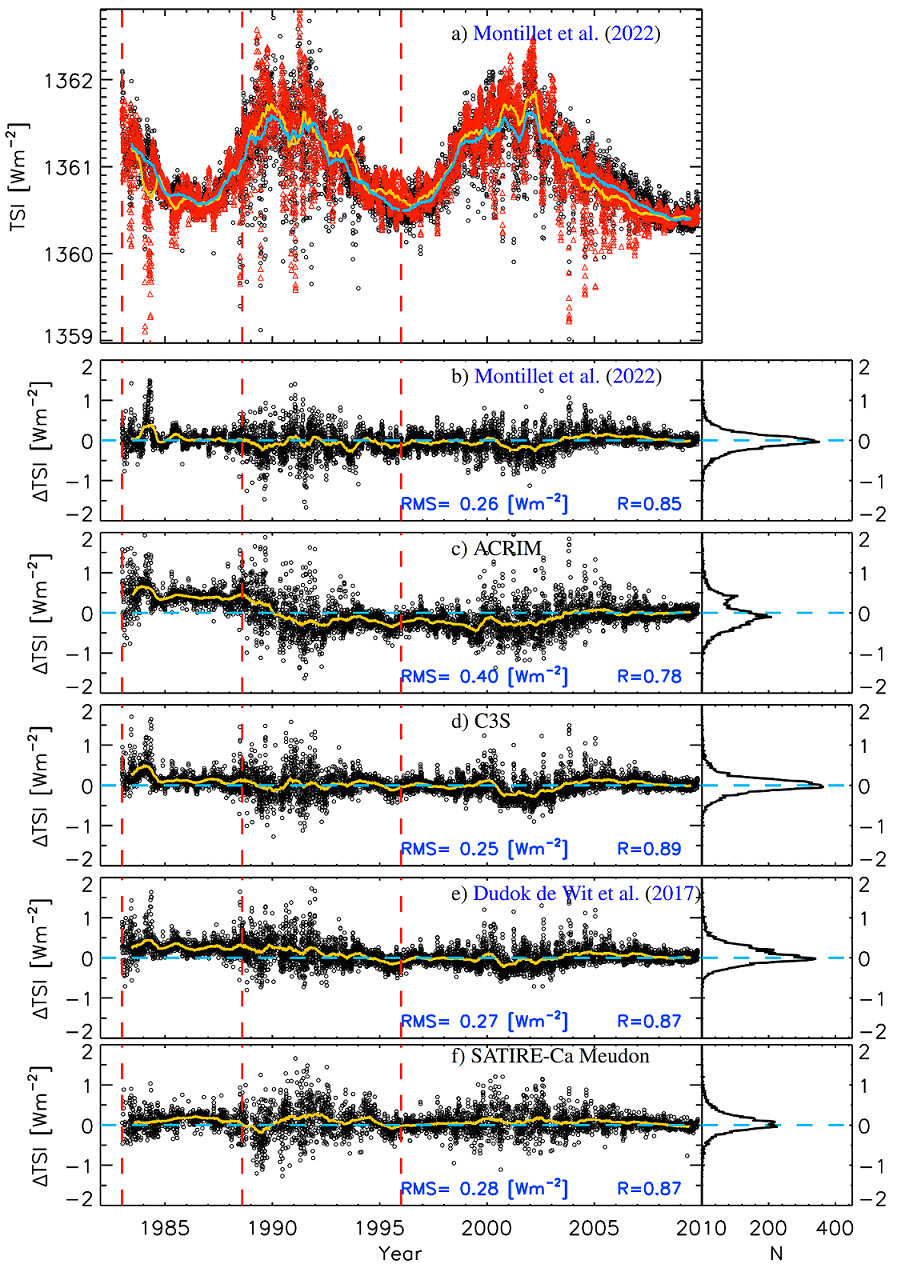}
	  \caption{Same as Fig. \ref{fig:tsirecmw2_unc} but now the Mt Wilson filling factors within the three distinct periods marked with the vertical dashed red lines have been separately calibrated to those from Meudon (until 1996) and SoHO/MDI (since 1996).
	  }\label{fig:tsirecmw2_merged}
\end{figure*}

\begin{figure*}[!htb]
	\centering
\includegraphics[width=0.8\linewidth]{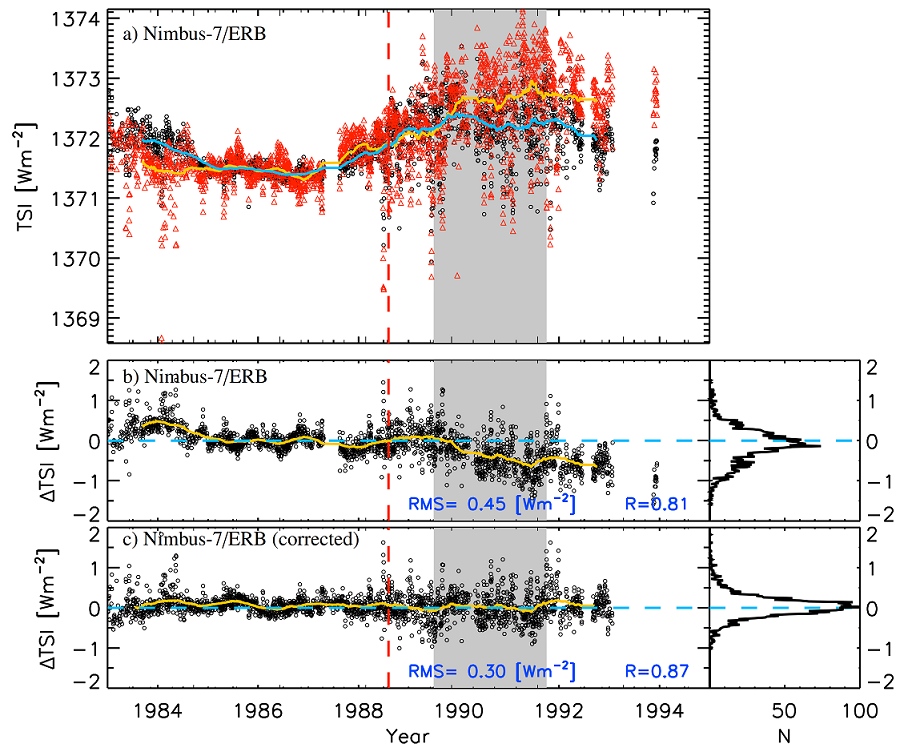}
	  \caption{\textit{panel a): }TSI reconstruction with the SATIRE-S model using Mt Wilson magnetograms (black circles for daily and ciel line for 180-day running means) along with the Nimbus-7/ERB direct TSI measurements (red triangles for daily and yellow line for 180-day running means). \textit{Panels b)--c): } 
   Comparison between TSI reconstruction with the SATIRE-S model using Mt Wilson magnetograms and direct TSI measurements over the ACRIM-gap (see legends in each panel). Shown are residuals (SATIRE-S with Mt Wilson minus the respective series in each panel) in daily values (black circles) and 180-day running means (yellow lines). The horizontal dashed ciel line marks residuals of 0 Wm$^{-2}$.  
   The vertical dashed red line marks the change in Mt Wilson instrumentation over August 1988. The grey shaded surface marks the ACRIM-gap period between July 1989 and October 1991. 
   The right panels show the histogram of the residuals in bins of 0.02 Wm$^{-2}$. 
	  }\label{fig:tsirecmw2}
\end{figure*}

\begin{figure*}[]
	\centering
\includegraphics[width=0.8\linewidth]{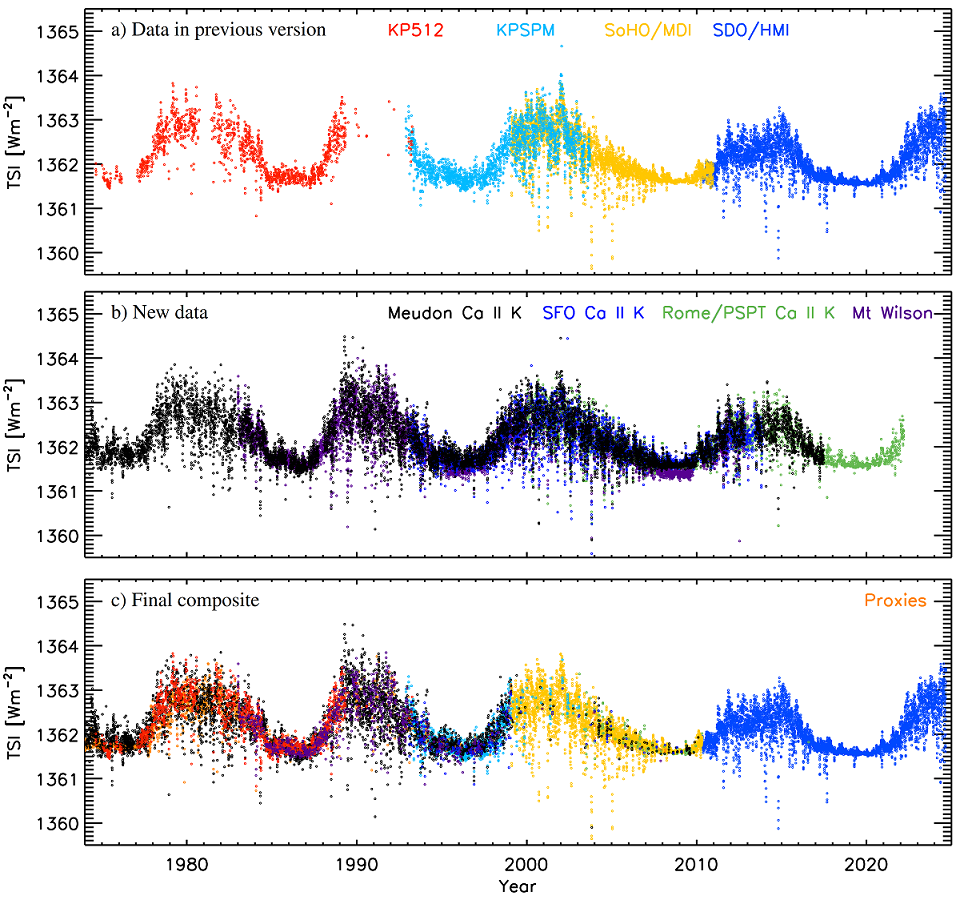}
 	  \caption{SATIRE-S TSI reconstruction using different sources of magnetograms (shown in different colours according to the legends). Shown are daily values. The data that were included in the \citet{yeo_reconstruction_2014} version of the SATIRE-S TSI composite are shown in panel a), while the new data considered in this work are shown in panel b).
    Panel c) shows the final new SATIRE-S TSI composite.
	  }\label{fig:updateSatire-s}
\end{figure*}

\begin{table}
\caption{Contrast threshold for umbra and penumbra identification.}
\label{tab:sunspotthresholds} 
\centering
\begin{tabular}{lcc}
\hline\hline
&$C_{p}$ & $C_{u}$\\
\hline                    
\centering
SDO/HMI		&	0.87 & 0.59 \\
SoHO/MDI	&	0.89 & 0.64 \\
KP/SPM		&	0.92 & 0.70 \\
Rome/PSPT	&	0.93 & 0.65 \\  
\hline
\end{tabular}
\end{table}

\begin{table*}
\caption{TSI differences between solar cycle minima.}
\label{tab:mintomintrends} 
\centering
\begin{tabular}{l*{6}{c}}
\hline\hline
TSI series&2019--1976&2019--1986&1986--1976&1996--1986&2008--1996&2019--2008\\			
          &SC 25--21 &SC 25--22 &SC 22--21 &SC 23--22 &SC 24--23 &SC 25--24\\	\hline
This work & -2.19 (1.72) & -0.58 (1.31) & -1.61 (2.11) &  0.19 (2.00) & -0.33 (1.58) & -0.43 (0.48) \\
\citet{yeo_reconstruction_2014} & -4.62 (0.91) & -4.59 (0.97) & -0.03 (1.26) & -2.19 (1.54) & -2.21 (1.27) & -0.20 (0.45) \\
\citet{montillet_data_2022} &  -  & -2.35 (1.16) &  -  &  0.09 (1.68) & -1.87 (1.34) & -0.57 (0.57) \\ 
ACRIM &  -  &  -  &  -  &  4.77 (2.09) & -3.08 (1.85) &  -  \\
ROB &  -  &  -  &  -  &  -  & -0.48 (1.25) &  0.88 (0.50) \\ C3S &  -  & -0.11 (1.25) &  -  &  0.13 (1.85) & -1.05 (1.47) &  0.81 (0.56) \\
\citet{dudok_de_wit_methodology_2017} &  -  &  -  &  -  &  1.76 (1.88) & -1.56 (1.47) &  -  \\
\hline
\end{tabular}
\tablefoot{The values are given as the absolute differences (outside parentheses) and uncertainty (inside parentheses) between the end and start years expressed in dW m$^{-2}$. Underneath the periods we also list the corresponding cycle numbers (where the years refer to the cycle start).
}
\end{table*}

\begin{figure*}[]
	\centering
\includegraphics[width=0.8\linewidth]{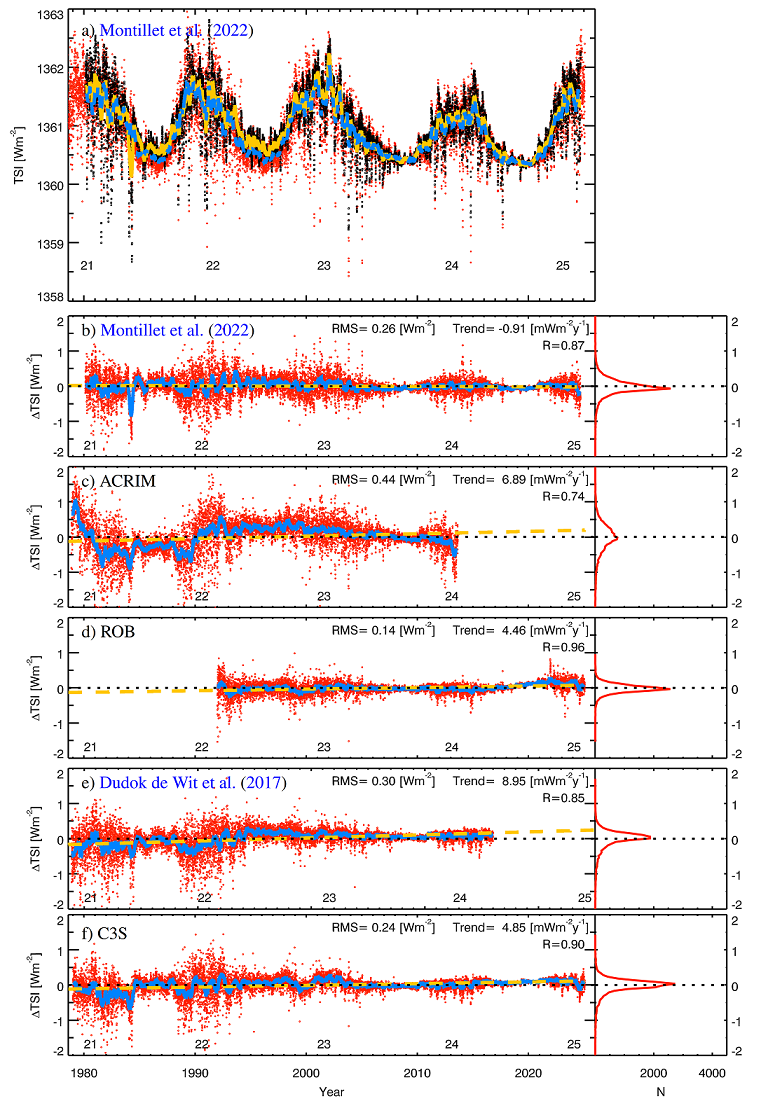}
 	  \caption{\textit{a) \ } The \citet{montillet_data_2022} TSI composite (black dots for daily values and thick yellow line for 180-day running means) overplotted on the updated SATIRE-S TSI composite (red dots for daily values and thick blue line for 180-day running means). \textit{b)--f) \ } Difference between various composites of measured TSI (as denoted in the legend) and the the updated SATIRE-S reconstructed TSI composite. Daily values are represented by red dots and 180-day running means by the thick blue line. 
    The dashed yellow lines are linear fits to the residuals. The RMS differences, linear correlation coefficients, and the trend of the residuals are given in each panel. The dotted black line marks residuals of 0 Wm$^{-2}$. The right panel shows the histogram of the residuals in bins of 0.05 Wm$^{-2}$.}
    \label{fig:updateSatire-s-measurements}
\end{figure*}

\begin{figure*}[]
	\centering
\includegraphics[width=0.8\linewidth]{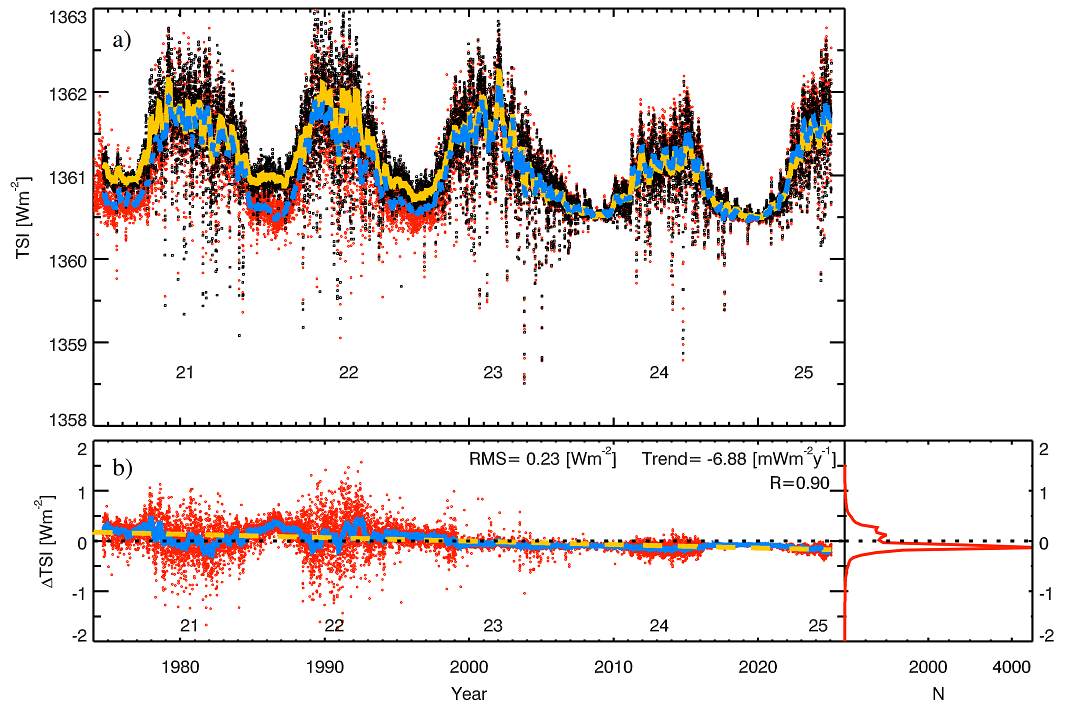}
    \caption{\textit{Top: } Updated SATIRE-S TSI composite (red dots for daily values and thick blue line for 180-day running means) in comparison to the earlier version by \citet{yeo_reconstruction_2014}  (black dots for daily values and thick yellow line for 180-day running means).     \textit{Bottom: } Difference between the \citet{yeo_reconstruction_2014} and updated versions. Daily values are given by the red dots and the 180-day running means by the blue line.  The dashed yellow line is a linear fit to the residuals. 
    The RMS difference, linear correlation coefficient, and the trend of the residuals are listed in panel b). The dotted black line marks residuals of 0 Wm$^{-2}$.  The right panel shows the histogram of the residuals in bins of 0.05 Wm$^{-2}$.
	  }\label{fig:updateSatire-s-oldSatire-s}
\end{figure*}

\begin{figure*}[]
	\centering
\includegraphics[width=0.9\linewidth]{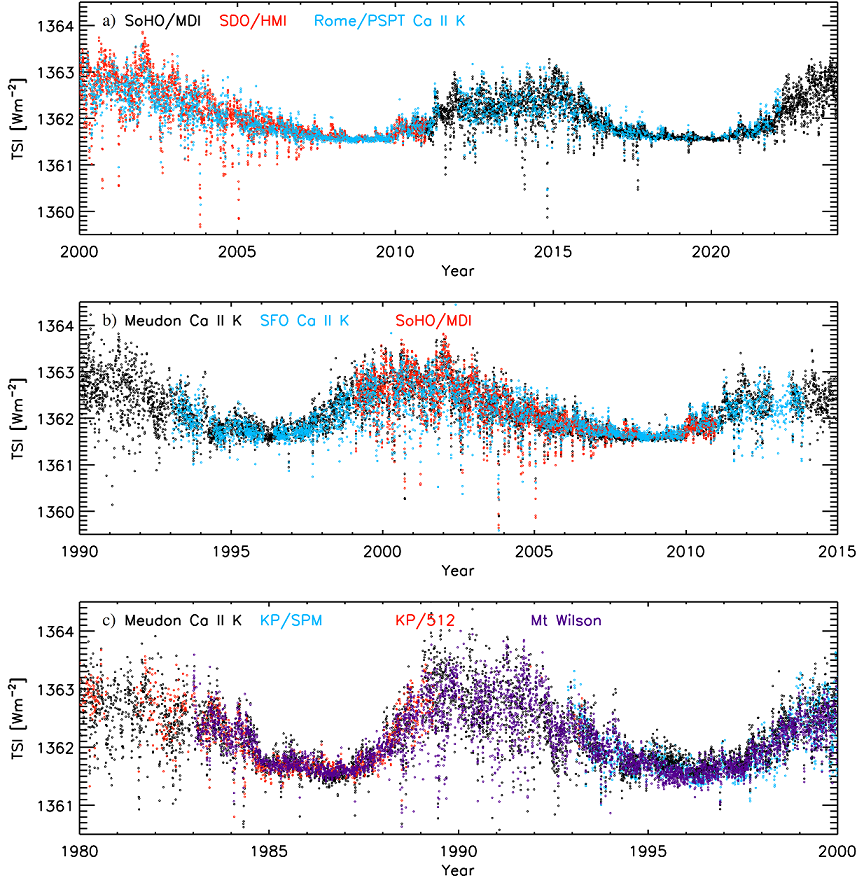}
 	  \caption{TSI reconstruction with SATIRE-S using different sets of magnetograms (shown in different colours according to the legends). The three panels focus on the minima of 2008 vs. 2019, 1996 vs. 2008, and 1986 vs. 1996, respectively.
	  }\label{fig:minima}
\end{figure*}

\begin{figure*}[]
	\centering
\includegraphics[width=1\linewidth]{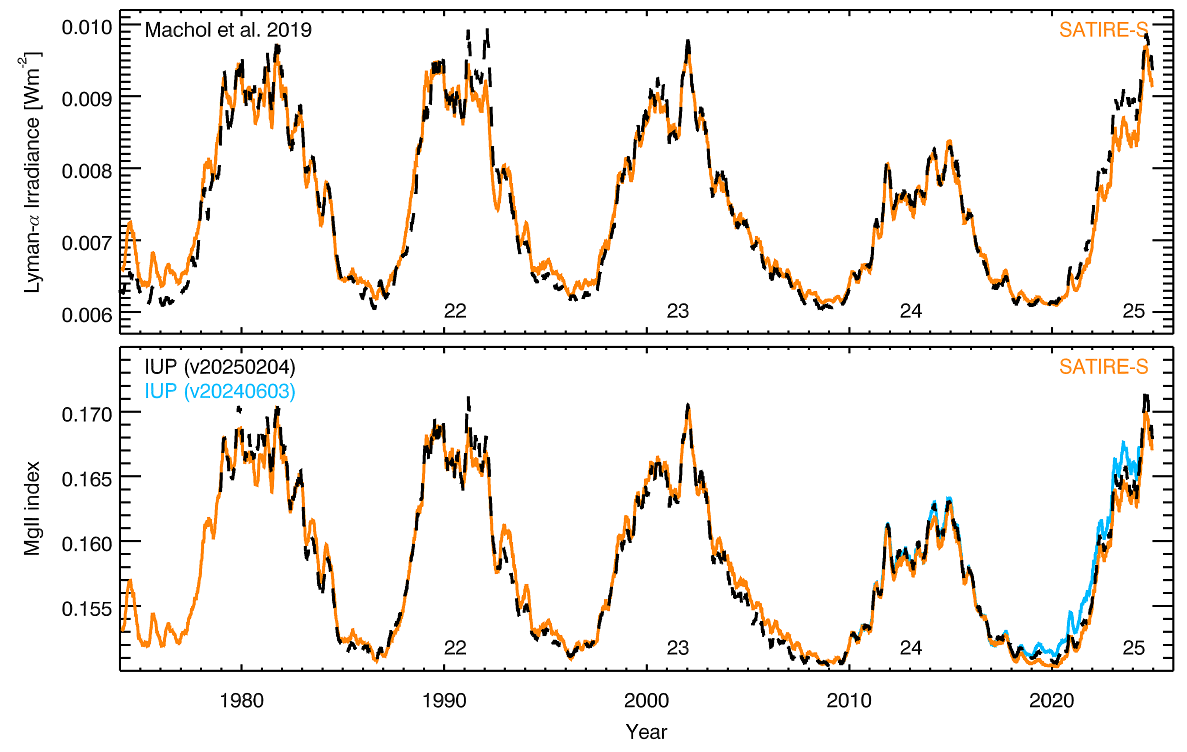}
 	  \caption{\textit{Top: }Comparison between the reconstructed Lyman-$\alpha$ irradiance from SATIRE-S (orange) and the \citet{machol_improved_2019} composite (black). Shown are 81-day running means. The series are shown without any adjustment. \textit{Bottom: }Comparison between the reconstructed Mg II irradiance from SATIRE-S (orange) and the IUP MgII core-to-wing composite \citep[][black for version 20250204 and ciel for version 20240603]{viereck_mg_2001}. Our Mg II irradiance has been linearly scaled to match the IUP Mg II index (see Sect. \ref{sec:compmeasurements}). The numbers in the lower part of the figures denote the conventional solar cycle numbering.
	  }\label{fig:updateSatire-s-measurementslya}
\end{figure*}

\section{Irradiance reconstruction}
\label{sec:tsirecs}
In this section we describe the overall reconstruction procedure, using  all the various sources of magnetograms.
Due to some intricacies with the Mt Wilson archive, its employment for the reconstruction is discussed separately in Sect.~\ref{sec:MWtsirecs}.
The final composite series is then described in Sect.~\ref{sec:newcomposite}.

\subsection{Filling factor derivation}
\label{sec:fillingfactors}
As discussed in Sect.~\ref{sec:satires}, the filling factors of faculae are derived from the magnetic flux density measured in the magnetograms, see Eq.~\ref{eq:bsat}.
We compute the filling factors as the fractional areas of faculae within 101 $\mu$ positions (the choice of the optimum $B_\mathrm{sat}$ is described in the next section, Sect.~\ref{sec:fixbsat}).

The identification of umbra and penumbra in Kitt Peak, SoHO/MDI, SDO/HMI, and Rome/PSPT datasets is done from near-co-temporal continuum observations from the same sites.
Umbra and penumbra were identified with constant contrast thresholds in the limb-darkening-corrected images (here the mean level of the quiet Sun is set to 1) listed in Table \ref{tab:sunspotthresholds}. 

Different approaches were used for the KP/512, Mt Wilson, Meudon, and SFO data.
The KP/512 continuum images have a limited dynamic range, making it difficult to accurately correct for the limb-darkening and thus also identify umbrae and penumbrae.
For this reason, \citet{wenzler_reconstruction_2006} used a different and more complex segmentation approach than a constant threshold to identify the locations of whole spots.
A constant ratio of 0.2 between umbra and whole spot area was used to derive the KP/512 umbra and penumbra filling factors.

The sunspot filling factors for Meudon, SFO, and Mt Wilson, for which we do not have continuum observations, were derived with the composite of foreshortening-corrected areas and locations of individual sunspot groups compiled by \citet{mandal_sunspot_2020} following \citet{chatzistergos_reconstructing_2021-1}.
\citet{chatzistergos_reconstructing_2021-1} used SATIRE to compute the TSI variations employing unsigned magnetic field maps restored from Rome/PSPT \ca observations in combination with either continuum observations or the database by \citet{mandal_sunspot_2020} to determine the sunspot filling factors.
Their results showed only a minute deterioration of the quality of the reconstruction when using the database by \citet{mandal_sunspot_2020} instead of co-temporal continuum observations.

To account for differences in  magnetograms from the various sources and to bring the filling factors to the same level, we followed the approach by \citet{yeo_reconstruction_2014}.
We used SDO/HMI as the main dataset and brought the filling factors of the other archives to the level of SDO/HMI using a daisy chain process.
The corresponding daisy-chaining sequence is given in Table \ref{tab:mwdatachains}.

\subsection{Fixing the free parameter}
\label{sec:fixbsat}
To set the free parameter, $B_\mathrm{sat}$, we compared our reconstructions to reference TSI measurements. 
As has been shown in previous studies \citep{krivova_acrim-gap_2009,yeo_reconstruction_2014,chatzistergos_reconstructing_2021-1,chatzistergos_understanding_2024}, the resulting reconstruction depends only weakly on the choice of the reference TSI record.
Nevertheless, to account for the potential uncertainty or bias related to the choice of the reference data set, we used all available  single-instrument  measurements  (see Sect. \ref{sec:tsimeas}) with a direct overlap to SDO/HMI data to set $B_\mathrm{sat}$, see Fig.~\ref{fig:hmi}, and thus to estimate its uncertainty range.
As optimum $B_\mathrm{sat}$, we consider the value that minimises the sum of the RMS differences between our reconstruction and the reference TSI series for daily and 180-day smoothed values.
We found $B_\mathrm{sat}$ values to lie between 255 G and 284 G for different TSI measurement records.
For our final reconstruction we adopt  the value of $B_\mathrm{sat}$=270~G, which is roughly in the middle of the determined  $B_\mathrm{sat}$ range.
This value lies closest to the one returned by a comparison to the TCTE/TIM instrument.
We use the extreme $B_\mathrm{sat}$ values to derive the uncertainty range of our reconstruction due to the choice of the reference data set, shown as yellow shaded surface in Fig.~\ref{fig:hmi}.

\subsection{UV correction}

The intensity spectra we use to compute irradiance (Eq.~\ref{eq:rec}), were synthesized assuming local thermodynamic equilibrium, which breaks down in the upper layers of the solar atmosphere, where the UV radiation forms.
This leads to a progressive divergence of our reconstruction from measurements below about 300 nm \citep[see][for details]{krivova_reconstruction_2006}.
From 180 to 300 nm, this is essentially an offset in the absolute levels of SSI, while between 115~nm and 180~nm, also the amplitude of the SSI variations is weaker.
Following the previous versions of SATIRE \citep{krivova_reconstruction_2006,yeo_reconstruction_2014}, we applied an empirical correction in this range.
From 180 to 300 nm we offset the SSI reconstruction to match the level of the Whole Heliospheric Interval (WHI) reference solar spectra \citep{woods_solar_2009}. 
From 115 to 180 nm we linearly scaled the computed SSI to observations by SORCE/SOLSTICE \citep{snow_solar-stellar_2022}.
This correction is time-independent and does not affect the long-term change, see \citet{yeo_reconstruction_2014} for further details.
This correction was independently validated by \citet{tagirov_readdressing_2019}, who used SATIRE with non-LTE intensity spectra and filling factors derived from SDO/HMI observations to compute the UV variability over the rising phase of cycle 24.

\subsection{Reconstructing irradiance with Mt Wilson magnetograms}
\label{sec:MWtsirecs}

As mentioned in Sect. \ref{sec:data_mtwilson}, there were several instrumental changes affecting the Mt Wilson dataset.
To address the archive's consistency we first performed an initial and preliminary reconstruction by considering all Mt Wilson magnetograms since 12 December 1981.
Only for this preliminary reconstruction do we use the filling factors without bringing them to the SDO/HMI level (see Sect. \ref{sec:fillingfactors}), thus we also used a different value for $B_\mathrm{sat}$=205 G. 
This reconstruction is shown in Fig. \ref{fig:tsirecmw2_unc} and it is compared to direct TSI measurements and our TSI reconstruction with Meudon \ca data (listed as SATIRE-ca Meudon).

In Fig. \ref{fig:tsirecmw2_unc}, we marked three dates when instrumental changes occurred: January 1983, August 1988, and January 1996. 
On all of these occasions, we observe similar trends when comparing to all TSI measurements, which suggests a link to the instrumental changes.  
If unaccounted, this would affect the long-term consistency of the reconstruction.
In particular, we observe distinct offsets in the reconstruction as compared to all TSI records considered here.
These are more pronounced during the first two periods, but a synchronous change in the residuals is also evident after the third change in 1996 in all panels. 
Potentially, there could be another inconsistency in Mt Wilson data over 2010 following a long data gap.
However, this is not important for our analysis considering that over that period there are high-quality SDO/HMI magnetograms and thus the Mt Wilson data over that period will not enter our final series.

We also noticed a sharp and short-lived change between our reconstruction with Mt Wilson data and the various TSI composites over 1984. 
However, it is absent in the comparison to our reconstruction with Meudon \ca data.
This suggests that the likely cause of this discrepancy is inconsistencies in the direct TSI measurements used by the composites and not issues with the Mt Wilson data.

The above analysis suggests that the Mt Wilson magnetograms cannot, unfortunately, be used as an autonomous homogeneous dataset over the entire period due the various instrumental issues and updates.
However, our analysis also demonstrates that these data can be used to accurately recover irradiance variations over individual periods when the instrument was running stably, i.e. between times when the instrument had issues or was changed.
Thus, after accounting for the inhomogeneities, the Mt Wilson data can be employed for filling gaps in the reconstructions with SATIRE-S from other sources of magnetograms.
To accurately account for the inhomogeneities, we consider the Mt Wilson data as three separate sets of data for which we cross-calibrate the filling factors individually (see Sect. \ref{sec:fillingfactors}).
The corresponding periods are January 1983 to July 1988, August 1988 to December 1995, and January 1996 to January 2013.
Although the first relevant instrumental change occurred in May 1982 we do not consider Mt Wilson data before 1983 to be on the conservative side.
The Mt Wilson filling factors over the first two intervals were cross-calibrated to Meudon \ca ones, while the third interval was cross-calibrated to SoHO/MDI  (we note that both Meudon and SoHO/MDI were first brought to the SDO/HMI level). 
We then used these cross-calibrated filling factors to reconstruct irradiance with $B_\mathrm{sat}$=270 G as derived for SDO/HMI magnetograms and subsequently  also used for all other magnetogram datasets considered here.

Figure \ref{fig:tsirecmw2_merged} shows TSI reconstructed with Mt Wilson data after cross-calibrating the filling factors over the three intervals separately.
This reconstruction is in a good agreement with the \citet{montillet_data_2022}, C3S, \citet{dudok_de_wit_methodology_2017}, and ROB TSI composites as well as the TSI SATIRE reconstruction using Meudon \ca data.
The agreement with ACRIM  TSI composite has improved for the period after 1991 compared to the original uncorrected version (Fig. \ref{fig:tsirecmw2_unc}), however there is a clearly different trend over 1989--1991.
This hints at residual issues in the data from Nimbus-7/ERB affecting this composite.
Figure~\ref{fig:tsirecmw2} shows a comparison of our reconstruction with Mt Wilson data (after cross-calibrating the filling factors) to Nimbus-7/ERB data.
We notice a clear discrepancy between our reconstruction and Nimbus-7/ERB TSI over 1990, with Nimbus-7/ERB TSI exhibiting an abrupt increase (best visible as the drop in the yellow curve around 1990 in Fig.~\ref{fig:tsirecmw2}b).
This increase is the same as also seen when comparing to the ACRIM TSI composite.
This has important implications for the so-called ACRIM-gap (covering July 1989 to October 1991).
That is because Mt Wilson is a completely independent dataset that covers the entire period of the ACRIM-gap and thus provides an important and independent check.
We find the same abrupt change when comparing Nimbus-7/ERB TSI data to our reconstruction with Meudon data, which is also a consistent and independent dataset.
Our results suggest that the increasing TSI in Nimbus-7/ERB data after 1990 is most likely due to an unaccounted instrumental issue of Nimbus-7/ERB.
Figure \ref{fig:tsirecmw2} also shows a comparison of our TSI reconstruction with Mt Wilson data to the Nimbus-7/ERB as they were corrected by \citet{frohlich_solar_2006}.
We find that our reconstruction is consistent with the corrections applied to Nimbus-7/ERB data. 
Thus, our analysis favours the overall corrections applied to Nimbus-7/ERB data by \citet{frohlich_solar_2006}.
We note that these corrections were not applied on the Nimbus-7/ERB data used in the ACRIM TSI composite, which leads to an artefact in its long-term behaviour, making this series less reliable.

\subsection{Updated SATIRE-S TSI and SSI composite}
\label{sec:newcomposite}

Here we combine the irradiance reconstructions from the different sources of the magnetograms presented in Sect.~\ref{sec:mag} to produce a revised and updated SATIRE-S TSI and SSI composite record.
Figure \ref{fig:updateSatire-s} shows our reconstructed TSI using the same $B_\mathrm{sat}=270$ G value after cross-calibrating the filling factors from each dataset in different colours.
We show in separate panels the reconstructions from the archives that were included in the previous version of the SATIRE-S TSI composite \citep[][panel a)]{yeo_reconstruction_2014} along with the new series we incorporated here (panel b)). 
For each day we keep data from only one source, giving highest priority to SDO/HMI and then progressively lower priority to the data from SoHO/MDI, KP/SPM, KP/512, Rome/PSPT, Meudon, SFO, and Mt Wilson.

Our set of magnetograms leaves 1771 data gaps ($\sim$9\% of the covered period), which we filled by regressing to our irradiance reconstruction from magnetograms derived from the five proxy indices (Sunspot areas, \ca plage areas, Lyman-$\alpha$, F10.7, and Mg II index) introduced in Sect. \ref{Sec:proxies}.
We use all five indices together in multi-linear regression. 
However, this still leaves gaps (1252 days) in the reconstruction, which we address by using three additional combinations of pairs of proxy series. 
These combination pair Lyman-$\alpha$ with sunspot areas, F10.7, and plage areas, respectively. 
With this we achieve a complete coverage since the 1st of January 1974.
Our final TSI composite is shown in Fig. \ref{fig:updateSatire-s}c) with various data sources marked in different colours.

\section{Comparison to measurements}
\label{sec:compmeasurements}
Figure \ref{fig:updateSatire-s-measurements} compares our final SATIRE-S TSI composite to direct measurements of TSI.
Overall, we find a very good agreement between our reconstruction and the various composites of TSI measurements, with linear correlation coefficients between 0.85 and 0.96 for daily values for all datasets except ACRIM.
The poorest agreement is with the ACRIM TSI composite, with a linear correlation coefficient of 0.74 and RMS difference of 0.44Wm$^{-2}$ for daily values.
A major contributor to that is the ACRIM-gap increase in TSI observed in the ACRIM composite, which appears to result from uncorrected issues in the Nimbus-7/ERB data as was used by ACRIM. 
Our TSI reconstructions using Mt Wilson and Meudon data, both consistent and of high quality during the ACRIM-gap period (see Sect. \ref{sec:MWtsirecs}), disfavour the TSI increase suggested by ACRIM.
Also, the long-term trend of TSI in the updated SATIRE-S composite agrees fairly well with most series. 
The residuals between our reconstruction and the \citet{montillet_data_2022} TSI composite exhibit a flat trend of -0.91 mW m$^{-2}$ y$^{-1}$.
For other TSI composites, the trends of the residuals range from 4.46 to 8.95 mW m$^{-2}$ y$^{-1}$, suggesting that
these records exhibit trends that are either less steeply decreasing or even increasing compared to our reconstruction.

Figure \ref{fig:updateSatire-s-oldSatire-s} compares the revised SATIRE-S TSI composite to the previous version by \citet{yeo_reconstruction_2014}.
Both versions exhibit rather similar characteristics, with minuscule residuals after 1999 and somewhat greater differences before that.
The RMS difference between the two series over the whole period is 0.24 Wm$^{-2}$ with a linear correlation coefficient of 0.89 for daily values.
Overall, the values at solar cycle maxima are comparable in the two versions, however, due to the changes in how we use SDO/HMI data (update of noise surface and revision of $B_\mathrm{sat}$, see Sects. \ref{sec:processing} and \ref{sec:fixbsat}), the amplitude of solar cycle 24 is slightly increased in the new version compared to the previous one, while the amplitude of cycle 22 is reduced in the new version (due to the increased value of $B_\mathrm{sat}$ compared to the previous version).
The most noticeable difference between the two series, however, is a slightly different overall long-term change.
That is, the previous version returned higher TSI values during the activity minima preceding solar cycles 22 and 23 compared to the new version, which translates into a minimum to minimum difference between 2019 and 1976 of $-0.22$ and $-0.46$ Wm$^{-2}$ for the revised and the \citet{yeo_reconstruction_2014} versions, respectively (see Table \ref{tab:mintomintrends}).
This is because of the significantly higher uncertainties over this period in the earlier version due to the very short overlap of the two KP magnetogram sets (see Fig. \ref{fig:updateSatire-s}).

Overall the trend in the updated SATIRE-S TSI composite is independent of measured TSI and relies on SDO/HMI, the connection between SoHO/MDI to SDO/HMI, and on the Meudon \ca archive for the period before 1996.
Although Meudon has consistent and high quality data over that period, there is a potential uncertainty in the calibration of their filling factors due to the instrumental changes over 2002, when they switched from using photographic plates to a CCD camera, as well as over 1989 when the spectroheliograph was upgraded \citep{malherbe_130_2023}. 
The very good agreement between the reconstructions using Meudon and SFO data (Figure \ref{fig:minima}b) verifies that the instrumental change over 2002 in Meudon does not bias the long-term trend of our reconstruction.

Based on the comparison between our reconstructions from Meudon and Mt Wilson data and the Nimbus-7/ERB TSI we argue that the instrumental change in Meudon over 1989 also does not affect its long-term trend.
Both reconstructions, based on Meudon and Mt Wilson data, align well with Nimbus-7/ERB after a -0.5 Wm$^{-2}$ offset is applied on Nimbus-7/ERB TSI data from 1990 onward (or if the Nimbus-7/ERB data corrected by \citealt{frohlich_solar_2006} are used). 
This suggests that the instrumental change in Meudon in January 1989 does not significantly impact our results.

We also note that the connection between SDO/HMI and SoHO/MDI is also further supported by the agreement to the reconstruction with Rome/PSPT data (Figure \ref{fig:minima}a).
Thus, with this reconstruction we are providing a more accurate estimate on the long-term trend in irradiance variations, which is derived entirely independently of the instrumental irradiance measurements.
The cycle minimum-to-minimum trends of our SATIRE-S TSI composite is given in Table \ref{tab:mintomintrends} for the various cycles and compared to those from composites of direct measurements.
We find a very small difference between the 1976 and 2019 minima of -0.22 $\pm 0.17$ Wm$^{-2}$, which is roughly half of the difference in the previous version \citep{yeo_reconstruction_2014}.
The difference between the minima in 1986 and 2019 in our reconstruction is in  very good agreement with that in the C3S composite ($-0.06 \pm 0.13$ and $-0.011 \pm 0.12$~Wm$^{-2}$ in our reconstruction and C3S, respectively), while \citet{montillet_data_2022} find a somewhat stronger downward trend of $-0.24 \pm 0.12$~Wm$^{-2}$.

We have updated both the SATIRE-S TSI and SSI reconstruction. However, we focused the discussion above on TSI.
This is because our revision aimed at the improvement of the estimate of the long-term trend, while the spectral distribution of the variability has not been affected by the current changes and remains as discussed by \citet{yeo_UV_2015} and \citet{tagirov_readdressing_2019}. 
For completeness, however, we briefly compare our reconstruction with two long-term series of UV irradiance variations, which are not affected by instrumental effects to the extent other SSI measurements are \citep[see][]{yeo_solar_2015}.
Figure \ref{fig:updateSatire-s-measurementslya} compares our Lyman-$\alpha$ (1\AA~spectral width) irradiance reconstruction to the \citet{machol_improved_2019} composite series.
We find an excellent agreement between the two series, reaching RMS differences of 3.2 $\times 10^{-4}$ Wm$^{-2}$ and a linear correlation coefficient of 0.96. 
Our reconstruction is a bit higher prior to 1978 than the \citet{machol_improved_2019} composite and slightly lower after 2022.
The agreement between our reonstruction and the \citet{machol_improved_2019} Lyman-$\alpha$ composite series is slightly better than for the previous version of SATIRE-S, for which the RMS difference is 3.7 $\times 10^{-4}$ Wm$^{-2}$, while the linear correlation coefficient is also 0.96. 

In Figure \ref{fig:updateSatire-s-measurementslya} we also compare our Mg II irradiance reconstruction (with 1\AA~spectral width) to the IUP Mg II composite index \citep{viereck_mg_2001}. 
Since the IUP index considers the core-to-wing ratio of the line, we linearly scaled our reconstruction to the IUP index so as to bring them to the same scale.
We find an excellent agreement between the two series, with a linear correlation coefficient of 0.97.

\section{Summary and conclusions}
\label{sec:summary}
Accurate knowledge of long-term variations in solar irradiance is essential for understanding the solar influence on Earth's climate. 
While direct regular measurements of total solar irradiance (TSI) have been available since 1978, they come from a series of relatively short-lived experiments, each with its own challenges, which do not allow a reliable estimate of the long-term changes.
Since solar surface magnetism drives irradiance fluctuations over periods longer than a day, models that use measurements of the solar photospheric magnetic field provide an alternative approach for assessing these variations. 

To enhance our understanding of the long-term irradiance changes during the period covered by direct measurements, we produced a new total and spectral irradiance (TSI and SSI) reconstruction with the SATIRE-S model.
SATIRE is a physics-based semi-empirical model which derives information on the evolution of solar surface magnetism from full-disc magnetograms of the Sun.
The previous version of SATIRE-S \citep{yeo_reconstruction_2014} used four datasets of magnetograms, SDO/HMI, SoHO/MDI, KP/SPM, and KP/512.
However, these archives covered roughly one solar cycle each with rather short overlaps between the various datasets which resulted in some uncertainties in their cross-calibration.
This meant that although TSI reconstructed by SATIRE-S was clearly decreasing over the period of satellite measurements, the exact magnitude of this decrease was comparatively uncertain.

In this work we incorporated four additional sources of magnetograms, namely the archive from Mt Wilson and the unsigned magnetograms reconstructed from three \ca archives  \citep[Meudon, Rome/PSPT, and San Fernando;][]{chatzistergos_recovering_2019,chatzistergos_reconstructing_2021-1}.
This allowed a TSI and SSI reconstruction with a significantly lower uncertainty in the long-term trend.  
The latter is entirely independent of the direct measurements of TSI and SSI.
The new SATIRE-S TSI and SSI composites will be available at \url{https://www2.mps.mpg.de/projects/sun-climate/data.html}.

Our new SATIRE-S TSI reconstruction shows excellent agreement with direct measurements of TSI.
The linear correlation coefficients between the reconstruction and TSIS1-TIM TSI is 0.98. The correlation coefficients with the composites by \citet{montillet_data_2022} and ROB are 0.87 and 0.96, respectively.
We also found an excellent agreement of the reconstructed Lyman-$ \alpha$ and Mg~II irradiance with the Lyman-$ \alpha$ \citet{machol_improved_2019} and the IUP Mg II index \citep{viereck_mg_2001} composites (with correlation coefficients of 0.96 and 0.97, respectively).

Our reconstruction exhibits a rather small change of TSI of $-0.22~\pm$ 0.17 Wm$^{-2}$ between the 1976 and 2019 minima, which is roughly half of that in the previous version  by \citet{yeo_reconstruction_2014}.
The difference between the 1986 and 2019 minima for our series, -0.06 W m$^{-2}$, is in good agreement to that of the C3S composite, but it is somewhat lower than the one from the \citet{montillet_data_2022} series, -0.24 W m$^{-2}$.

Our analysis also provides an independent way of reconciling the ACRIM-gap \citep{krivova_acrim-gap_2009}.
That is because two of the new sources of magnetograms, Mt Wilson and Meudon, cover the ACRIM-gap completely.
Both Meudon and Mt Wilson reconstructions disfavour the increase in TSI between 1989 and 1991, while our composite reconstruction exhibits a minuscule, and statistically insignificant, increase of 0.02 $\pm$ 0.2 Wm$^{-2}$ between the 1986 and 1996 minima.
This is in stark contrast to the 0.48 Wm$^{-2}$ increase reported by the ACRIM TSI composite.
By comparing the Nimbus7-ERB TSI data to our reconstructions we argue for residual issues in the Nimbus7-ERB series manifesting as an abrupt increase of TSI of about 0.5 Wm$^{-2}$ over 1990, which would account completely for the ACRIM-gap.
This increase affects the ACRIM TSI composite since it relies on the uncorrected Nimbus7-ERB data over that period.

In our reconstruction we filled gaps for about 9\% of days for which we do not have magnetograms by regressing proxies, while we restricted the use of magnetograms to data starting from 1974. 
Other sources of magnetograms such as those from Global Oscillation Network Group \citep[GONG;][]{leibacher_global_1999} or the Vector Spectromagnetograph (VSM) on the Synoptic Optical Long-term Investigations of the Sun (SOLIS) system \citep{keller_solis_2003} as well as reconstructed unsigned magnetograms from other archives of \ca observations can be incorporated to fill further gaps in the reconstruction and reduce the usage of the regressed proxies as well as extend our reconstruction back to 1892.
However, substantial effort is still required to address various issues with the \ca data and to evaluate their consistency \citep{chatzistergos_understanding_2024}.

\begin{acknowledgements}
We thank the anonymous reviewer for their feedback.
This work was supported by the German Federal Ministry of Education and Research (Project No. 01LG1909C) and by the
European Union's Horizon 2020 research and Innovation program under grant agreement No 824135 (SOLARNET).
This study includes data from the synoptic program at the 150-Foot Solar Tower of the Mt. Wilson Observatory. 
The Mt. Wilson 150-Foot Solar Tower is operated by UCLA, with funding from NASA, ONR and NSF, under agreement with the Mt. Wilson Institute.
This project has received funding from the European Research Council (ERC) under the European Union's Horizon 2020 research and innovation programme (grant agreement No. 101097844 — project WINSUN).
This study has made use of SAO/NASA Astrophysics Data System's bibliographic services.
\end{acknowledgements}

\bibliographystyle{aa}
\bibliography{biblio01}

\begin{thebibliography}{82}
\expandafter\ifx\csname natexlab\endcsname\relax\def\natexlab#1{#1}\fi

\bibitem[{Babcock \& Babcock(1955)}]{babcock_suns_1955}
Babcock, H.~W. \& Babcock, H.~D. 1955, The Astrophysical Journal, 121, 349

\bibitem[{Ball {et~al.}(2012)Ball, Unruh, Krivova, Solanki, Wenzler, Mortlock,
  \& Jaffe}]{ball_reconstruction_2012}
Ball, W.~T., Unruh, Y.~C., Krivova, N.~A., {et~al.} 2012, Astronomy and
  Astrophysics, 541, A27

\bibitem[{Chapman {et~al.}(2024)Chapman, Cookson, \&
  Choudhary}]{chapman_tsi_2024}
Chapman, G.~A., Cookson, A.~M., \& Choudhary, D.~P. 2024, Journal of Space
  Weather and Space Climate, 14, 34, publisher: EDP Sciences

\bibitem[{Chapman {et~al.}(2013)Chapman, Cookson, \&
  Preminger}]{chapman_modeling_2013}
Chapman, G.~A., Cookson, A.~M., \& Preminger, D.~G. 2013, Solar Physics, 283,
  295

\bibitem[{Chatzistergos {et~al.}(2020{\natexlab{a}})Chatzistergos, Ermolli,
  Giorgi, Krivova, \& Puiu}]{chatzistergos_modelling_2020}
Chatzistergos, T., Ermolli, I., Giorgi, F., Krivova, N.~A., \& Puiu, C.~C.
  2020{\natexlab{a}}, Journal of Space Weather and Space Climate, 10, 45

\bibitem[{Chatzistergos {et~al.}(2019{\natexlab{a}})Chatzistergos, Ermolli,
  Krivova, \& Solanki}]{chatzistergos_analysis_2019}
Chatzistergos, T., Ermolli, I., Krivova, N.~A., \& Solanki, S.~K.
  2019{\natexlab{a}}, Astronomy \& Astrophysics, 625, A69

\bibitem[{Chatzistergos {et~al.}(2020{\natexlab{b}})Chatzistergos, Ermolli,
  Krivova, Solanki, Banerjee, Barata, Belik, Gafeira, Garcia, Hanaoka, Hegde,
  Klimeš, Korokhin, Lourenço, Malherbe, Marchenko, Peixinho, Sakurai, \&
  Tlatov}]{chatzistergos_analysis_2020}
Chatzistergos, T., Ermolli, I., Krivova, N.~A., {et~al.} 2020{\natexlab{b}},
  Astronomy \& Astrophysics, 639, A88

\bibitem[{Chatzistergos {et~al.}(2018)Chatzistergos, Ermolli, Solanki, \&
  Krivova}]{chatzistergos_analysis_2018}
Chatzistergos, T., Ermolli, I., Solanki, S.~K., \& Krivova, N.~A. 2018,
  Astronomy \& Astrophysics, 609, A92

\bibitem[{Chatzistergos {et~al.}(2019{\natexlab{b}})Chatzistergos, {Ermolli,
  Ilaria}, {Solanki, Sami K.}, {Krivova, Natalie A.}, {Giorgi, Fabrizio}, \&
  {Yeo, Kok Leng}}]{chatzistergos_recovering_2019}
Chatzistergos, T., {Ermolli, Ilaria}, {Solanki, Sami K.}, {et~al.}
  2019{\natexlab{b}}, Astronomy \& Astrophysics, 626, A114

\bibitem[{Chatzistergos {et~al.}(2022)Chatzistergos, Krivova, \&
  Ermolli}]{chatzistergos_full-disc_2022}
Chatzistergos, T., Krivova, N.~A., \& Ermolli, I. 2022, Frontiers in Astronomy
  and Space Sciences, 9

\bibitem[{Chatzistergos {et~al.}(2024)Chatzistergos, Krivova, \&
  Ermolli}]{chatzistergos_understanding_2024}
Chatzistergos, T., Krivova, N.~A., \& Ermolli, I. 2024, Journal of Space
  Weather and Space Climate, 14, 9

\bibitem[{Chatzistergos {et~al.}(2021)Chatzistergos, Krivova, Ermolli, Yeo,
  Mandal, Solanki, Kopp, \& Malherbe}]{chatzistergos_reconstructing_2021-1}
Chatzistergos, T., Krivova, N.~A., Ermolli, I., {et~al.} 2021, Astronomy \&
  Astrophysics, 656, A104

\bibitem[{Chatzistergos {et~al.}(2023)Chatzistergos, Krivova, \&
  Yeo}]{chatzistergos_long-term_2023}
Chatzistergos, T., Krivova, N.~A., \& Yeo, K.~L. 2023, Journal of Atmospheric
  and Solar-Terrestrial Physics, 252, 106150

\bibitem[{Dewitte {et~al.}(2004)Dewitte, Crommelynck, \&
  Joukoff}]{dewitte_total_2004}
Dewitte, S., Crommelynck, D., \& Joukoff, A. 2004, Journal of Geophysical
  Research (Space Physics), 109, A02102

\bibitem[{Dewitte \& Nevens(2016)}]{dewitte_total_2016}
Dewitte, S. \& Nevens, S. 2016, The Astrophysical Journal, 830, 25

\bibitem[{Domingo {et~al.}(1995)Domingo, Fleck, \& Poland}]{domingo_soho_1995}
Domingo, V., Fleck, B., \& Poland, A.~I. 1995, Solar Physics, 162, 1

\bibitem[{Dudok~de Wit {et~al.}(2017)Dudok~de Wit, Kopp, Fröhlich, \&
  Schöll}]{dudok_de_wit_methodology_2017}
Dudok~de Wit, T., Kopp, G., Fröhlich, C., \& Schöll, M. 2017, Geophysical
  Research Letters, 44, 1196

\bibitem[{Ermolli {et~al.}(2022)Ermolli, Giorgi, \&
  Chatzistergos}]{ermolli_rome_2022}
Ermolli, I., Giorgi, F., \& Chatzistergos, T. 2022, Frontiers in Astronomy and
  Space Sciences, 9

\bibitem[{Fligge {et~al.}(2000{\natexlab{a}})Fligge, Solanki, \&
  Unruh}]{fligge_modelling_2000-1}
Fligge, M., Solanki, S.~K., \& Unruh, Y.~C. 2000{\natexlab{a}}, Astronomy and
  Astrophysics, 353, 380

\bibitem[{Fligge {et~al.}(2000{\natexlab{b}})Fligge, Solanki, \&
  Unruh}]{fligge_modelling_2000}
Fligge, M., Solanki, S.~K., \& Unruh, Y.~C. 2000{\natexlab{b}}, Space Science
  Reviews, 94, 139

\bibitem[{Fontenla {et~al.}(1999)Fontenla, White, Fox, Avrett, \&
  Kurucz}]{fontenla_calculation_1999}
Fontenla, J., White, O.~R., Fox, P.~A., Avrett, E.~H., \& Kurucz, R.~L. 1999,
  The Astrophysical Journal, 518, 480

\bibitem[{Fröhlich(2000)}]{frohlich_observations_2000}
Fröhlich, C. 2000, Space Science Reviews, 94, 15

\bibitem[{Fröhlich(2006)}]{frohlich_solar_2006}
Fröhlich, C. 2006, Space Science Reviews, 125, 53

\bibitem[{Fröhlich(2012)}]{frohlich_total_2012}
Fröhlich, C. 2012, Surveys in Geophysics, 33, 453

\bibitem[{Fröhlich {et~al.}(1995)Fröhlich, Romero, Roth, Wehrli, Andersen,
  Appourchaux, Domingo, Telljohann, Berthomieu, Delache, Provost, Toutain,
  Crommelynck, Chevalier, Fichot, Däppen, Gough, Hoeksema, Jiménez, Gómez,
  Herreros, Roca~Cortés, Jones, Pap, \& Willson}]{frohlich_virgo_1995}
Fröhlich, C., Romero, J., Roth, H., {et~al.} 1995, Solar Physics, 162, 101

\bibitem[{Gray {et~al.}(2010)Gray, Beer, Geller, Haigh, Lockwood, Matthes,
  Cubasch, Fleitmann, Harrison, Hood, Luterbacher, Meehl, Shindell, van Geel,
  \& White}]{gray_solar_2010}
Gray, L.~J., Beer, J., Geller, M., {et~al.} 2010, Reviews of Geophysics, 48,
  4001

\bibitem[{Haigh(2007)}]{haigh_sun_2007}
Haigh, J.~D. 2007, Living Reviews in Solar Physics, 4, 2

\bibitem[{Howard {et~al.}(1983)Howard, Boyden, Bruning, Clark, Crist, \&
  Labonte}]{howard_mount_1983}
Howard, R., Boyden, J.~E., Bruning, D.~H., {et~al.} 1983, Solar Physics, 87,
  195

\bibitem[{Hoyt {et~al.}(1992)Hoyt, Kyle, Hickey, \&
  Maschhoff}]{hoyt_nimbus_1992}
Hoyt, D.~V., Kyle, H.~L., Hickey, J.~R., \& Maschhoff, R.~H. 1992, Journal of
  Geophysical Research, 97, 51

\bibitem[{{IPCC}(2021)}]{masson-delmotte_climate_2021}
{IPCC}. 2021, Climate {Change} 2021: {The} {Physical} {Science} {Basis}.
  {Contribution} of {Working} {Group} {I} to the {Sixth} {Assessment} {Report}
  of the {Intergovernmental} {Panel} on {Climate} {Change}, ed.
  V.~Masson-Delmotte, P.~Zhai, A.~Pirani, S.~Connors, C.~Péan, S.~Berger,
  N.~Caud, Y.~Chen, L.~Goldfarb, M.~Gomis, M.~Huang, K.~Leitzell, E.~Lonnoy,
  J.~Matthews, T.~Maycock, T.~Waterfield, O.~Yelekçi, R.~Yu, \& B.~Zhou
  (Cambridge, United Kingdom and New York, NY, USA: Cambridge University Press)

\bibitem[{Jones \& Ceja(2001)}]{jones_preliminary_2001}
Jones, H.~P. \& Ceja, J.~A. 2001, in {ASP} {Conference} {Proceedings}, Vol. 236
  (San Francisco: Astronomical Society of the Pacific), 87

\bibitem[{Kahil {et~al.}(2017)Kahil, Riethmüller, \&
  Solanki}]{kahil_brightness_2017}
Kahil, F., Riethmüller, T.~L., \& Solanki, S.~K. 2017, The Astrophysical
  Journal Supplement Series, 229, 12, number: 1

\bibitem[{Keller {et~al.}(2003)Keller, Harvey, \& Giampapa}]{keller_solis_2003}
Keller, C.~U., Harvey, J.~W., \& Giampapa, M.~S. 2003, in Innovative
  {Telescopes} and {Instrumentation} for {Solar} {Astrophysics}, Vol. 4853
  (SPIE), 194--204

\bibitem[{Kopp(2016)}]{kopp_magnitudes_2016}
Kopp, G. 2016, Journal of Space Weather and Space Climate, 6, A30

\bibitem[{Kopp {et~al.}(2013)Kopp, Boyle, Heuerman, Pilewskie, Seidel, Swieter,
  \& Ucker}]{kopp_tcte_2013}
Kopp, G., Boyle, B., Heuerman, K., {et~al.} 2013, AGU Fall Meeting Abstracts,
  51, GC51C

\bibitem[{Kopp {et~al.}(2005)Kopp, Lawrence, \& Rottman}]{kopp_total_2005}
Kopp, G., Lawrence, G., \& Rottman, G. 2005, in The {Solar} {Radiation} and
  {Climate} {Experiment} ({SORCE}), ed. G.~Rottman, T.~Woods, \& V.~George
  (Springer New York), 129--139

\bibitem[{Kren {et~al.}(2017)Kren, Pilewskie, \& Coddington}]{kren_where_2017}
Kren, A.~C., Pilewskie, P., \& Coddington, O. 2017, Journal of Space Weather
  and Space Climate, 7, A10

\bibitem[{Krivova(2018)}]{krivova_solar_2018}
Krivova, N.~A. 2018, in Climate {Changes} in the {Holocene} (CRC Press),
  107--120

\bibitem[{Krivova {et~al.}(2003)Krivova, Solanki, Fligge, \&
  Unruh}]{krivova_reconstruction_2003}
Krivova, N.~A., Solanki, S.~K., Fligge, M., \& Unruh, Y.~C. 2003, Astronomy and
  Astrophysics, 399, L1

\bibitem[{Krivova {et~al.}(2006)Krivova, Solanki, \&
  Floyd}]{krivova_reconstruction_2006}
Krivova, N.~A., Solanki, S.~K., \& Floyd, L. 2006, Astronomy and Astrophysics,
  452, 631

\bibitem[{Krivova {et~al.}(2009)Krivova, Solanki, \&
  Wenzler}]{krivova_acrim-gap_2009}
Krivova, N.~A., Solanki, S.~K., \& Wenzler, T. 2009, Geophysical Research
  Letters, 36, L20101

\bibitem[{Kurucz(1970)}]{kurucz_atlas_1970}
Kurucz, R.~L. 1970, SAO Special Report, 309

\bibitem[{Kurucz(1993)}]{kurucz_new_1993}
Kurucz, R.~L. 1993, in International {Astronomical} {Union}, {Colloquium} {No}.
  138, held in {Trieste}, {Italy}, {July} 1992, Vol.~44, Astronomical {Society}
  of the {Pacific}, ed. M.~dworetsky, F.~Castelli, \& R.~Faraggiana (San
  Francisco: Astronomical Society of the Pacific), 87, conference Name: IAU
  Colloq. 138: Peculiar versus Normal Phenomena in A-type and Related Stars

\bibitem[{Kurucz(2005)}]{kurucz_atlas12_2005}
Kurucz, R.~L. 2005, Memorie della Societa Astronomica Italiana Supplementi, 8,
  14

\bibitem[{Leibacher(1999)}]{leibacher_global_1999}
Leibacher, J.~W. 1999, Advances in Space Research, 24, 173

\bibitem[{Liu {et~al.}(2012)Liu, Hoeksema, Scherrer, Schou, Couvidat, Bush,
  Duvall, Hayashi, Sun, \& Zhao}]{liu_comparison_2012}
Liu, Y., Hoeksema, J.~T., Scherrer, P.~H., {et~al.} 2012, Solar Physics, 279,
  295

\bibitem[{Loukitcheva {et~al.}(2009)Loukitcheva, Solanki, \&
  White}]{loukitcheva_relationship_2009}
Loukitcheva, M., Solanki, S.~K., \& White, S.~M. 2009, Astronomy and
  Astrophysics, 497, 273

\bibitem[{Machol {et~al.}(2019)Machol, Snow, Woodraska, Woods, Viereck, \&
  Coddington}]{machol_improved_2019}
Machol, J., Snow, M., Woodraska, D., {et~al.} 2019, Earth and Space Science, 6,
  2263

\bibitem[{Malherbe(2023)}]{malherbe_130_2023}
Malherbe, J.-M. 2023, Journal for the History of Astronomy, 54, 274

\bibitem[{Malherbe \& Dalmasse(2019)}]{malherbe_new_2019}
Malherbe, J.-M. \& Dalmasse, K. 2019, Solar Physics, 294, 52

\bibitem[{Mandal {et~al.}(2020)Mandal, Krivova, Solanki, Sinha, \&
  Banerjee}]{mandal_sunspot_2020}
Mandal, S., Krivova, N.~A., Solanki, S.~K., Sinha, N., \& Banerjee, D. 2020,
  Astronomy \& Astrophysics, 640, A78

\bibitem[{Montillet {et~al.}(2022)Montillet, Finsterle, Kermarrec, Sikonja,
  Haberreiter, Schmutz, \& Dudok~de Wit}]{montillet_data_2022}
Montillet, J.-P., Finsterle, W., Kermarrec, G., {et~al.} 2022, Journal of
  Geophysical Research: Atmospheres, 127, e2021JD036146

\bibitem[{Pesnell {et~al.}(2012)Pesnell, Thompson, \&
  Chamberlin}]{pesnell_solar_2012}
Pesnell, W.~D., Thompson, B.~J., \& Chamberlin, P.~C. 2012, Solar Physics, 275,
  3

\bibitem[{Pilewskie {et~al.}(2018)Pilewskie, Kopp, Richard, Coddington, Sparn,
  \& Woods}]{pilewskie_tsis-1_2018}
Pilewskie, P., Kopp, G., Richard, E., {et~al.} 2018, in Proceedings from the
  20th {EGU} {General} {Assembly} held 4-13 {April}, 2018 in {Vienna},
  {Austria}, Vol.~20, 5527

\bibitem[{Scherrer {et~al.}(1995)Scherrer, Bogart, Bush, Hoeksema, Kosovichev,
  Schou, Rosenberg, Springer, Tarbell, Title, Wolfson, Zayer, \& {MDI
  Engineering Team}}]{scherrer_solar_1995}
Scherrer, P.~H., Bogart, R.~S., Bush, R.~I., {et~al.} 1995, Solar Physics, 162,
  129

\bibitem[{Scherrer {et~al.}(2012)Scherrer, Schou, Bush, Kosovichev, Bogart,
  Hoeksema, Liu, Duvall, Zhao, Title, Schrijver, Tarbell, \&
  Tomczyk}]{scherrer_helioseismic_2012}
Scherrer, P.~H., Schou, J., Bush, R.~I., {et~al.} 2012, Solar Physics, 275, 207

\bibitem[{Schmutz(2021)}]{schmutz_changes_2021}
Schmutz, W.~K. 2021, Journal of Space Weather and Space Climate, 11, 40

\bibitem[{Schrijver {et~al.}(1989)Schrijver, Cote, Zwaan, \&
  Saar}]{schrijver_relations_1989}
Schrijver, C.~J., Cote, J., Zwaan, C., \& Saar, S.~H. 1989, The Astrophysical
  Journal, 337, 964

\bibitem[{Shapiro {et~al.}(2017)Shapiro, Solanki, Krivova, Cameron, Yeo, \&
  Schmutz}]{shapiro_nature_2017}
Shapiro, A.~I., Solanki, S.~K., Krivova, N.~A., {et~al.} 2017, Nature
  Astronomy, 1, 612

\bibitem[{Skumanich {et~al.}(1975)Skumanich, Smythe, \&
  Frazier}]{skumanich_statistical_1975}
Skumanich, A., Smythe, C., \& Frazier, E.~N. 1975, The Astrophysical Journal,
  200, 747

\bibitem[{Snow {et~al.}(2022)Snow, McClintock, Woods, \&
  Elliott}]{snow_solar-stellar_2022}
Snow, M., McClintock, W.~E., Woods, T.~N., \& Elliott, J.~P. 2022, Solar
  Physics, 297, 55

\bibitem[{Solanki {et~al.}(2013)Solanki, Krivova, \&
  Haigh}]{solanki_solar_2013-1}
Solanki, S.~K., Krivova, N.~A., \& Haigh, J.~D. 2013, Annual Review of
  Astronomy and Astrophysics, 51, 311

\bibitem[{Solanki \& Stenflo(1984)}]{solanki_properties_1984}
Solanki, S.~K. \& Stenflo, J.~O. 1984, Astronomy and Astrophysics, 140, 185

\bibitem[{Tagirov {et~al.}(2019)Tagirov, Shapiro, Krivova, Unruh, Yeo, \&
  Solanki}]{tagirov_readdressing_2019}
Tagirov, R.~V., Shapiro, A.~I., Krivova, N.~A., {et~al.} 2019, Astronomy and
  Astrophysics, 631, A178

\bibitem[{Tapping(2013)}]{tapping_107_2013}
Tapping, K.~F. 2013, Space Weather, 11, 394

\bibitem[{Ulrich(1992)}]{ulrich_analysis_1992}
Ulrich, R.~K. 1992, in {ASP} conference series, ed. M.~S. Giampapa \& J.~A.
  Bookbinder, Vol.~26, 265

\bibitem[{Ulrich {et~al.}(2024)Ulrich, Boyden, \&
  Tran}]{ulrich_calibration_2024}
Ulrich, R.~K., Boyden, J., \& Tran, T. 2024, Solar Physics, 299, 149

\bibitem[{Ulrich \& Boyden(2019)}]{ulrich_calibration_2019}
Ulrich, R.~K. \& Boyden, J.~E. 2019, in {AGU} {Fall} {Meeting}, Vol. 2019,
  SH41B--01, abstract \#SH41B-01

\bibitem[{Ulrich {et~al.}(2002)Ulrich, Evans, Boyden, \&
  Webster}]{ulrich_mount_2002}
Ulrich, R.~K., Evans, S., Boyden, J.~E., \& Webster, L. 2002, The Astrophysical
  Journal Supplement Series, 139, 259

\bibitem[{Ulrich {et~al.}(1991)Ulrich, Webster, Boyden, Magnone, \&
  Bogart}]{ulrich_system_1991}
Ulrich, R.~K., Webster, L., Boyden, J.~E., Magnone, N., \& Bogart, R.~S. 1991,
  Solar Physics, 135, 211

\bibitem[{Unruh {et~al.}(1999)Unruh, Solanki, \& Fligge}]{unruh_spectral_1999}
Unruh, Y.~C., Solanki, S.~K., \& Fligge, M. 1999, Astronomy and Astrophysics,
  345, 635

\bibitem[{Viereck {et~al.}(2001)Viereck, Puga, McMullin, Judge, Weber, \&
  Tobiska}]{viereck_mg_2001}
Viereck, R., Puga, L., McMullin, D., {et~al.} 2001, Geophysical Research
  Letters, 28, 1343

\bibitem[{Wenzler {et~al.}(2006)Wenzler, Solanki, Krivova, \&
  Fröhlich}]{wenzler_reconstruction_2006}
Wenzler, T., Solanki, S.~K., Krivova, N.~A., \& Fröhlich, C. 2006, Astronomy
  and Astrophysics, 460, 583

\bibitem[{Willson(1997)}]{willson_total_1997}
Willson, R.~C. 1997, Science, 277, 1963

\bibitem[{Willson \& Mordvinov(2003)}]{willson_composite_2003}
Willson, R.~C. \& Mordvinov, A.~V. 2003, AGU Fall Meeting Abstracts, 31

\bibitem[{Woods {et~al.}(2009)Woods, Chamberlin, Harder, Hock, Snow, Eparvier,
  Fontenla, McClintock, \& Richard}]{woods_solar_2009}
Woods, T.~N., Chamberlin, P.~C., Harder, J.~W., {et~al.} 2009, Geophysical
  Research Letters, 36, L01101, number: 1

\bibitem[{Yeo {et~al.}(2015{\natexlab{a}})Yeo, Ball, Krivova, Solanki, Unruh,
  \& Morrill}]{yeo_UV_2015}
Yeo, K.~L., Ball, W.~T., Krivova, N.~A., {et~al.} 2015{\natexlab{a}}, Journal
  of Geophysical Research (Space Physics), 120, 6055

\bibitem[{Yeo {et~al.}(2015{\natexlab{b}})Yeo, Krivova, \&
  Solanki}]{yeo_solar_2015}
Yeo, K.~L., Krivova, N.~A., \& Solanki, S.~K. 2015{\natexlab{b}}, in Space
  {Sciences} {Series} of {ISSI}, Vol.~53, The {Solar} {Activity} {Cycle}
  (Springer New York), 137

\bibitem[{Yeo {et~al.}(2014)Yeo, Krivova, Solanki, \&
  Glassmeier}]{yeo_reconstruction_2014}
Yeo, K.~L., Krivova, N.~A., Solanki, S.~K., \& Glassmeier, K.~H. 2014,
  Astronomy and Astrophysics, 570, A85

\bibitem[{Yeo {et~al.}(2013)Yeo, Solanki, \& Krivova}]{yeo_intensity_2013}
Yeo, K.~L., Solanki, S.~K., \& Krivova, N.~A. 2013, Astronomy and Astrophysics,
  550, 95

\bibitem[{Yeo {et~al.}(2024)Yeo, Solanki, \& Krivova}]{yeo_variation_2024}
Yeo, K.~L., Solanki, S.~K., \& Krivova, N.~A. 2024, Astronomy \& Astrophysics,
  688, A48

\bibitem[{Yeo {et~al.}(2017)Yeo, Solanki, Norris, Beeck, Unruh, \&
  Krivova}]{yeo_solar_2017}
Yeo, K.~L., Solanki, S.~K., Norris, C.~M., {et~al.} 2017, Physical Review
  Letters, 119

\end{thebibliography}
\end{document}